\documentclass[pra,superscriptaddress,aps]{revtex4}

\usepackage{amsmath,epsfig,amsfonts,graphicx,hyperref}
\usepackage{subfigure}
\usepackage{graphicx}
\usepackage[american]{babel}
\usepackage{amssymb}
\usepackage{amsfonts}
\usepackage{color}
\usepackage[normalem]{ulem}
\begin{document}
	\newcommand{\be}{\begin{equation}}
	\newcommand{\ee}{\end{equation}}
	\newtheorem{corollary}{Corollary}[section]
	\newtheorem{remark}{Remark}[section]
	\newtheorem{definition}{Definition}[section]
	\newtheorem{theorem}{Theorem}[section]
	\newtheorem{proposition}{Proposition}[section]
	\newtheorem{lemma}{Lemma}[section]
	\newtheorem{help1}{Example}[section]

\def\re{\text{Re}}
\def\im{\text{Im}}
\def\labelitemi{$\blacktriangleright$}

\title{The linearly damped nonlinear Schr\"odinger equation with localized driving: spatiotemporal decay estimates and the emergence of extreme wave events}

\author{G. Fotopoulos}
\affiliation{Department of Electrical Engineering  and Computer Science,\\ Khalifa University P.O. Box 127788,  Abu Dhabi, United Arab Emirates}
\affiliation{Department of Mathematics, Statistics and Physics,\\ College of Arts and
	Sciences, Qatar University, P.\,O. Box 2713, Doha, Qatar}
\author{N.\,I. Karachalios}
\affiliation{Department of Mathematics,\\ Laboratory of Applied Mathematics and Mathematical Modelling,\\ University of the Aegean, Karlovassi, 83200
	Samos, Greece}
\author{V. Koukouloyannis}
\affiliation{Department of Mathematics,\\
Laboratory of Applied Mathematics and Mathematical Modelling,\\ University of the Aegean, Karlovassi, 83200
	Samos, Greece}
\affiliation{Department of Mathematics, Statistics and Physics,\\ College of Arts and
	Sciences, Qatar University, P.\,O. Box 2713, Doha, Qatar}
\author{K. Vetas}
\affiliation{Department of Mathematics, Statistics and Physics,\\ College of Arts and
	Sciences, Qatar University, P.\,O. Box 2713, Doha, Qatar}

\begin{abstract}
We prove spatiotemporal algebraically decaying estimates for the density of the solutions of  the linearly damped  nonlinear Schr\"odinger equation with localized driving, when supplemented with vanishing boundary conditions. Their derivation is made via a scheme, which incorporates suitable weighted Sobolev spaces and a time-weighted energy method. Numerical simulations examining the dynamics (in the presence of physically relevant examples of driver types and driving amplitude/linear loss regimes), showcase that the suggested decaying rates, are proved relevant in describing the transient dynamics of the solutions, prior their decay: they support the emergence of waveforms possessing an algebraic space-time localization (reminiscent of the Peregrine soliton) as first events of the dynamics, but also effectively capture the space-time asymptotics of the numerical solutions.
\end{abstract}

\maketitle

\section{Introduction}
In our recent works \cite{LetArx1,LetArx2}, we demonstrated, mainly by direct numerical simulations, the emergence of spatiotemporally localized wave forms reminiscent of the Peregrine rogue wave (PRW) \cite{H_Peregrine}, for the 
linearly damped and driven Schr\"{o}dinger (NLS) equation
\begin{eqnarray}
\label{eq1}
\mathrm{i}{{u}_{t}}+\frac{1}{2} {{u}_{xx}}+|u|^2u =f-\mathrm{i}\gamma u,\;\;\gamma\geq 0,
\end{eqnarray}
excited by  initial conditions $u(x,0)=u_0(x)$ vanishing at infinity, i.e.,
\begin{equation}\label{eq2Let}	
\lim_{|x|\rightarrow \infty}u_0(x)=0.
\end{equation}
In \cite{LetArx1}, we considered the case of  a Gaussian driver $f=f(x,t)$,  and for the problem supplemented with vanishing boundary conditions, we found that the PRW-structures emerged on the top of a decaying support. Particularly, we found that pending on the spatial/temporal scales of the driver, the transient dynamics – prior the ultimate decay of the solutions, resembles  the  one  in  the  semiclassical  limit  of  the  integrable  NLS \cite{CLM}, \cite{BM1}, \cite{BM2} (or  may lead  to  large-amplitude breather-like patterns). On the other hand, in \cite{LetArx2}, we considered the case of a spatially homogeneous (time-periodic) driver. In this case, we found that the PRW-type events are formed on the top of a self-induced finite background. This  dynamics was justified in terms of the existence and modulational instability of spatially homogeneous solutions of the model, and the continuous dependence of the solutions on the localized initial data for
small time intervals. 

In the present study, we aim to move a step forward to the analytical considerations for the problem discussed in \cite{LetArx1}: For equation \eqref{eq1}, when endowed with vanishing boundary  and initial conditions
\begin{equation}\label{eq2}	
\lim_{|x|\rightarrow \infty} u(x,t)=0,\;\;\mbox{for all}\;\;t\geq 0,
\end{equation}
we prove spatiotemporal decaying estimates for its solutions. These estimates assume the spatiotemporal localization of the driver
\begin{eqnarray}
\label{locdri}
\lim_{|x|^2+|t|^2\rightarrow\infty}f(x,t)=0,
\end{eqnarray}
at suitable rates.  Although under condition \eqref{locdri}, the global attractor \cite{Ghid88},\cite{XW95},\cite{Goubet1},\cite{Goubet2},\cite{Goubet3},\cite{Goubet2a},\cite{Lauren95},\cite{NK2002} for the associated dynamical system is trivial, the transient dynamics are highly non-trivial, as described above; the analytical estimates are found to be strongly relevant to the transient behavior of the system and  to the emergence of extreme wave events, as it will be explained in what follows.

Our approach in studying the spatially decaying estimates of the solutions is motivated by the algebraic spatiotemporal localization of the PRW. Thus, we  study the dynamics of Eq.~(\ref{eq1}), for initial data, possessing at least, an algebraic decaying rate. For the forcing term,  the assumption \eqref{locdri} is guided by an elementary physical intuition, in order to model external forces,  which are acting on a restricted spatial area and for limited time. With such assumptions, we revisit the well known functional-analytic approach whose origins are traced back in the seminal work \cite{Bab90}: that is, to derive energy estimates in suitable weighted Sobolev spaces whose elements are the considered initial data. At this point however, we should note that the lines of approach we shall follow herein, differ from the relevant ones, concerning evolution equations incorporating higher order dissipation, e.g., as the complex Ginzburg--Landau equation \cite{Mie1},\cite{Mie2}.  First, being interested in spatially localized solutions as the PRW--rather than fronts, periodic or quasi-periodic solutions--the selected weight should be unbounded instead of being integrable in the real line. Second, additional implications are caused by the absence of higher order dissipation, which may impose further restrictions to the chosen weight (e.g. on the boundedness of its derivative); these implications may be enhanced, by the necessity to derive some uniform in time estimates (as guided by the numerical results of \cite{LetArx1}). Third, although  the unbounded weights have been introduced in \cite{Bab90} (for the study of global attractors of localized solutions for reaction-difussion equations), or have been used at an auxiliary step to deduce the compactness of the attractor for the NLS \eqref{eq1} in standard Sobolev spaces (through a suitable decomposition of the semigroup)  \cite{Lauren95},\cite{NK2002}, here, we aim to highlight the usefulness of the approach in deriving point-wise estimates for the density of its solutions; up to our knowledge this approach is novel itself in order to study the spatial asymptotics of the solutions of Eq. \eqref{eq1} in the presence of the localized driver.

Specifically, the weighted energy estimates, combined with standard interpolation inequalities, allow for the derivation of  pointwise estimates for the spatial decay of the density of the solutions,  uniformly for all $t>0$. The implementation of the above procedure relies on the following key points: First, one should avoid a radiation boundary condition imposed by the spatial-weight. For this purpose, we prove the existence of solutions satisfying the prescribed spatial estimates through an approximation of the original problem in the real line. This approximation is defined by a sequence of problems considered in finite intervals $[-L,L]$ supplemented with Dirichlet boundary conditions \cite{Bab90},\cite{NK1999}. Second, the passage to the limit as $L\rightarrow\infty$,  is crucially based on the independence of the various constants of the aforementioned interpolation inequalities on $L$, for functions satisfying the vanishing boundary conditions (\ref{eq2}). Such an independence is not generically valid in the case of other boundary conditions, e.g., periodic or Neumann; we refer to \cite{ibc1,ibc2,ibc3} for  independent (best) constants for periodic functions with zero mean. Also, of major importance for the passage to the limit (in the nonlinear terms), are the compact inclusions between the involved functional spaces. 

Next, guided by the remarkable feature of the PRW, its algebraic temporal localization,  we proceed to the derivation of  time-decay estimates, which are uniform on the spatial variable $x$. We implement a time-weighted energy method in the spirit of \cite{HS}, under a suitable condition on the space-time weighted integrability of the forcing term. This condition suggests a certain time-decay rate for the forcing, in addition to its  required spatial decay rate. 
Yet the independence on the spatial domain of the various constants mentioned above, allow for the passage, from norm to point-wise time decay estimates, for the density of the solutions.

Several technical conditions raised in the process of the analytical arguments, which are connected to the primary question on the emergence of spatiotemporal localized solutions. The considered examples of algebraic weights, suggest at least, space-time algebraically decaying rates,  for  some specific types of localized drivers $f(x,t)$: the strongly localized one, assuming the functional form of a Gaussian function studied in \cite{LetArx1}, and a weakly localized driver possessing algebraic decaying rates. 

With such examples for an external driver, the damped and driven NLS \eqref{eq1} may become potentially relevant in the context of weakly nonlinear waves in the presence of wind forcing \cite{Kharif1},\cite{Kharif2},\cite{EPeli},\cite{onorato},\cite{brunetti},\cite{ChabchoubN}: Note that in this context, the damped and driven NLS incorporates linear dissipation with the damping strength being associated to the fluid's (e.g., water's) viscosity, but also linear forcing (competitive to damping) associated to the effect of wind forcing. Herein, the considered cases of drivers may serve as simplified phenomenological examples representing the effect of a spatiotemporal wave amplification due to the external forcing $f$ (inducing ``wind'' effects); we also refer to the recent work \cite{DHZ19} assuming the presence of a stochastic linear driving. 

In this regard, the usefulness of the analytical decaying estimates to a potential extreme wave dynamics is underlined by the findings of a numerical study.  For the numerical experiments, we have considered as in \cite{LetArx1},\cite{LetArx2}, a physically justified regime of forcing amplitude $\Gamma\sim O(1)$ and damping strength $\gamma\sim O(10^{-2})$. Such a regime could describe the impact of strong external  forcing $\Gamma$, against small fluid viscosity $\gamma$ (see \cite{brunetti}, \cite{ChabchoubN}).  Recalling some of the numerical results of \cite{LetArx1} for the Gaussian driver, we proceed to the study of the dynamics in the presence of the weakly localized forcing. In the present study, we consider both cases, of algebraically and exponentially decaying initial conditions \eqref{eq2Let}. 
For both examples of the driving forces and decaying initial conditions, we found that the  numerical spatiotemporal decaying rates are not only covered by the analytical estimates, but they also support  the emergence of PRW-algebraically localized structures as first events in the dynamics at the initial stages of the evolution. In addition, in most of the cases, these estimates are in excellent agreement with the analytical decaying rates, rationalizing the far space-time asymptotics observed numerically. 

We conclude the numerical explorations by examining the undamped, driven limit $\gamma=0$, $\Gamma>0$; we remark that this limit is not captured by our analytical arguments. On the one hand, the emergence of a PRW-type event yet persists in this limit. On the other hand, afterwards, we observed a totally different behaviour compared to the damped and forced case: we observe a focusing phenomenon around a breather-like mode of increasing amplitude. The dynamics shows that the solution becomes unbounded, as the forcing inserts energy continuously, although spatiotemporally localized. 

The paper is structured as follows. In Section~\ref{SII}, we present
the analytical considerations.
In Section~\ref{SIII}, we report the results of the numerical simulations. Finally,
in Section~\ref{SIV}, we summarize our results and briefly describe future directions.

\section{Generic spatiotemporal decay-estimates}
\label{SII}
In this section, we present the analytical considerations on the derivation of generic spatiotemporal decay estimates, for the solutions of the problem (\ref{eq1})-(\ref{eq2}). These estimates depend on the rate of decay of the initial conditions, as well as, on the decaying rates of the forcing term. More precisely, the decay rates of the initial data and of the driver are determined by the assumption, that both are elements of suitable weighted normed spaces. The section is divided in two subsections. In the subsection \ref{SUSA}, we derive spatial decaying estimates, while in  \ref{SUSB}, we derive the relevant time-decaying ones. 
\subsection{Spatial decay estimates}
\label{SUSA}
The derivation of spatially decaying estimates has two main ingredients. The first is the implementation of well known, uniform in time bounds in the standard Sobolev spaces \cite{Ghid88},\cite{XW95},\cite{Goubet1},\cite{Goubet2},\cite{Goubet3},\cite{Goubet2a},\cite{Lauren95},\cite{NK2002}. The second is the derivation of estimates in weighted normed spaces, for the solution and its spatial derivatives. For this purpose,  in order to avoid any radiation condition as $|x|\rightarrow\infty$ (and thus, any a-priori restriction on the spatial decaying rates of the solutions), we follow an approximation scheme similar to that of \cite{Bab90,NK1999}: at a first step,  we derive uniform-in time estimates on  weighted normed spaces for an auxiliary initial-boundary value problem supplemented with Dirichlet boundary conditions
\begin{equation}\label{eq2aux}	
u(-L,t)=u(L,t)=0,\;\;\mbox{for all}\;\;t\geq 0.
\end{equation}
It is crucial that the above weighted-normed estimates are also uniform on $L$. Thus, by the auxiliary problem, we may define an approximating sequence of solutions $u^{L}$. At a second step, we show that the approximating solutions $u^{L}$, converge to a solution $u$ of the problem (\ref{eq1})-(\ref{eq2}) as $L\rightarrow\infty$, which is at least of the class $u\in C[(0,\infty), H^1(\mathbb{R})]$, and uniformly bounded in the considered weighted normed spaces. Then, the derivation of pointwise decay estimates follows by a suitable application of standard interpolation inequalities. 

The first ingredient, the estimates in usual Sobolev spaces, is provided by the following theorem:
\begin{theorem}
\label{L1}  Let $u_0\in H^2_0(\mathcal{Q})$,  where either  $\mathcal{Q}=\mathbb{R}$ and $u$ is the solution of the problem (\ref{eq1})-(\ref{eq2}), or $\mathcal{Q}=(-L,L)$, and  $u$ is the solutions of the problem (\ref{eq1})-(\ref{eq2aux}).  We assume that $f,\,f_t\in L_{\mathrm{loc}}^{\infty}(\mathbb{R},L^2(\mathcal{Q}))$. Then, there exists a numerical constant $R$ which depends on $u_0,f,f_t$ and $\gamma$, and is independent of $t$ and $L$, such that the unique solution $u\in C([0,\infty),H^2_0(\mathcal{Q}))$ with $u_t\in C([0,\infty),L^2(\mathcal{Q}))$, satisfies the uniform in time estimate
\begin{eqnarray}
\label{eq4}
\|u\|_{H^2_0}^2\leq R.
\end{eqnarray}
\end{theorem}
Let us recall for notational purposes, that when $\mathcal{Q}=\mathbb{R}$, $H^2_0(\mathcal{Q})\equiv H^2(\mathbb{R})$. 

The second ingredient involves the following weighted-normed spaces, \cite{NK2002,Bab90}; we consider a weight function $\rho(x)$ with the following properties:
\begin{eqnarray}
\label{eqn5}
\rho(x)>0,\;\;\forall\; x\in\mathbb{R},\;\;\lim_{|x|\rightarrow\infty}\rho(x)=+\infty,\;\;0<\beta_1<|\rho'(x)|<\beta_2,\;\;\forall\;x\in\mathbb{R},\;\;\beta_1,\beta_2=\mathrm{const.}
\end{eqnarray}
Considering weight functions with the properties (\ref{eqn5}), the weighted-$L^2$ space, is defined as
$$
L^2_{\rho^2}(\mathbb{R}):=\left\{u:\mathbb{R}\rightarrow\mathbb{C}\;\;:\;\;
	\|u\|_{0,\rho^2}^2=\int_{\mathbb{R}}\rho^2|u|^2dx<\infty\right\},
$$
endowed with the real inner product $\left<u,v\right>_{\rho^2}=\mathrm{Re}\int_{\mathbb{R}}\rho^2u\bar{v}dx$, for all $u,v\in L^2_{\rho^2}$. Clearly, $L^2_{\rho^2}(\mathbb{R})\subset L^2(\mathbb{R})$, and consequently $L^p_\mathrm{loc}(\mathbb{R}, L^2_{\rho^2}(\mathbb{R}))\subset L^p_\mathrm{loc}(\mathbb{R}, L^2(\mathbb{R}))$, for all $1\leq p\leq\infty$, the inclusions being continuous, \cite{BrezisFA,ZeiIIA}. Accordingly, we shall use the weighted Sobolev space:
$$
H^1_{\rho^2}(\mathbb{R}):=\left\{u:\mathbb{R}\rightarrow\mathbb{C}\;\;:\;\;\|u\|_{1,\rho^2}^2=
\int_{\mathbb{R}}\rho^2|u|^2dx+\int_{\mathbb{R}}\rho^2|u_x|^2dx<\infty\right\},
$$
to deal with the first-order derivative of the solutions. In the following lemmas, we prove the estimates for the solutions of the auxiliary problem (\ref{eq1})-(\ref{eq2aux}), in the above weighted norms.
\begin{lemma}
	\label{LemW1}
Consider the Dirichlet initial-boundary value problem (\ref{eq1})-(\ref{eq2aux}) on $\mathcal{Q}=(-L,L)$. Let $u_0\in  H^2_0(\mathcal{Q})\cap L^2_{\rho^2}(\mathcal{Q})$, and $f\in L^{\infty}(\mathbb{R},L^2_{\rho^2}(\mathcal{Q}))$. Then, its solution is uniformly bounded in $L^2_{\rho^2}(\mathcal{Q})$: there exists a constant $R_0$ independent of $t$ and $L$, such that
	\begin{eqnarray}
		\label{eqn8}
		\|u\|^2_{0,\rho^2}< R_0,\;\;\forall\;\; t\in [0,\infty).
	\end{eqnarray}
\end{lemma}
\textbf{Proof:} We multiply (\ref{eq1}) by $\rho^2(x)\bar{u}$, and then integrate over $\mathcal{Q}$, keeping  the imaginary parts. The weighted $L^2$-norm balance equation is
\begin{eqnarray}
	\label{eqn9}
	\frac{1}{2}\frac{d}{dt}\int_{\mathcal{Q}}\rho^2|u|^2dx
	+\gamma\int_{\mathcal{Q}}\rho^2|u|^2dx+\frac{1}{2}\mathrm{Im}\int_{\mathcal{Q}}\rho^2u_{xx}\bar{u}dx=\mathrm{Im}\int_{\mathcal{Q}}\rho^2f\bar{u}dx.
\end{eqnarray}
For the third term of the left-hand side, we observe that:
\begin{eqnarray}
	\label{eqn10}
	\frac{1}{2}\mathrm{Im}\int_{\mathcal{Q}}\rho^2u_{xx}\bar{u}dx&=&\frac{1}{2}\left[\rho^2u_x\bar{u}\right]_{-L}^{L}-\frac{1}{2}\mathrm{Im}\int_{\mathcal{Q}}u_x(\rho^2\bar{u})_xdx\nonumber\\
	&=&-\mathrm{Im}\int_{\mathcal{Q}}\rho\rho'u_x\bar{u}dx-\frac{1}{2}\mathrm{Im}\int_{\mathcal{Q}}\rho^2u_x\bar{u}_xdx\nonumber\\
	&=&-\mathrm{Im}\int_{\mathcal{Q}}\rho\rho'u_x\bar{u}dx.
\end{eqnarray}
Let us remark from (\ref{eqn10}), that the weighted-$L^2$-energy consideration, already generates  a higher-order derivative term (the integral involving $u_x$ in our case). Although there is absence of higher dissipation, this term can be estimated with the help of Theorem \ref{L1}: by using the estimate (\ref{eq4}), it follows that  $\int_{\mathcal{Q}}|u_x|^2dx<R$. Hence, by using (\ref{eqn5}), 
\begin{eqnarray}
	\label{eqn11}
	\frac{1}{2}\left|\mathrm{Im}\int_{\mathcal{Q}}\rho^2u_{xx}\bar{u}dx\right|\leq \beta_2\int_{\mathcal{Q}}\rho|u_x\bar{u}|dx&\leq& \frac{\gamma}{4}\int_{\mathcal{Q}}\rho^2|u|^2dx+\frac{\beta_2^2}{\gamma}\int_{\mathcal{Q}}|u_x|^2dx\nonumber\\
	&\leq&\frac{\gamma}{4}\int_{\mathcal{Q}}\rho^2|u|^2dx+\frac{R\beta_2^2}{\gamma}.
\end{eqnarray}
Similarly, for the integral involving the driver, we have:
\begin{eqnarray}
\label{eqn12}
\left|\mathrm{Im}\int_{\mathcal{Q}}\rho^2f\bar{u}dx\right|\leq \frac{\gamma}{4}\int_{\mathcal{Q}}\rho^2|u|^2dx+\frac{1}{\gamma}\int_{\mathcal{Q}}\rho^2|f|^2dx.
\end{eqnarray}
Then, inserting (\ref{eqn11}) and (\ref{eqn12}) into (\ref{eqn9}), the latter becomes
\begin{eqnarray}
\label{eqn13}
\frac{1}{2}\frac{d}{dt}\int_{\mathcal{Q}}\rho^2|u|^2dx+\frac{\gamma}{2}\int_{\mathcal{Q}}\rho^2|u|^2dx\leq \frac{1}{\gamma}\left(R\beta_2^2+\int_{\mathcal{Q}}\rho^2|f|^2dx\right).
\end{eqnarray}
Taking the supremum for $t\in [0,\infty)$, on the right-hand side of \eqref{eqn13}, and applying Gronwall's lemma, we derive the existence of a numerical constant $M_0$, depending  on $\gamma$, on the norms of $u_0$, and $f$ in $L^{\infty}(\mathbb{R},L^2_{\rho^2}(\mathcal{Q}))$, but not on $t$ and $L$, such that
\begin{eqnarray}
\label{eqn14}
	\|u(t)\|_{0,\rho^2}^2\leq 	\|u_0\|_{0,\rho^2}^2\exp(-\gamma t)+M_0(1-\exp(-\gamma t)).
\end{eqnarray}
From the inequality (\ref{eqn14}), the estimate (\ref{eqn8}) readily follows.\ \  $\Box$  
\begin{lemma}
	\label{LemW2} Consider the Dirichlet initial-boundary value problem (\ref{eq1})-(\ref{eq2aux}) on $\mathcal{Q}=(-L,L)$. Let $u_0\in  H^2_0(\mathcal{Q})\cap H^1_{\rho^2}(\mathcal{Q})$, and $f,\,f_t\in L^{\infty}(\mathbb{R}^+,L^2_{\rho^2}(\mathcal{Q}))$. Then, its solution is uniformly bounded in $H^1_{\rho^2}(\mathcal{Q})$: there exists a constant $R_1$ independent of $t$ and $L$, such that
	\begin{eqnarray}
	\label{eqn14es}
	\|u\|^2_{1,\rho^2}< R_1,\;\;\forall\;\; t\in [0,\infty).
	\end{eqnarray}
\end{lemma}
\textbf{Proof:} This time, we multiply Eq. (\ref{eq1}) by $-\rho^2\bar{u}_t$, and integrate over $\mathcal{Q}$, keeping the real part of the resulting equation. Let us note that
\begin{eqnarray*}
-\mathrm{Re}\int_{\mathcal{Q}}\rho^2\bar{u}_t|u|^2udx=-\int_{\mathcal{Q}}\rho^2|u|^2\mathrm{Re}(\bar{u}_t u)dx=-\frac{1}{2}\int_{\mathcal{Q}}\rho^2|u|^2\frac{d}{dt}|u|^2dx,
\end{eqnarray*}
and since $\frac{d}{dt}|u|^4=2|u|^2\frac{d}{dt}|u|^2$, we have that
\begin{eqnarray*}
	-\mathrm{Re}\int_{\mathcal{Q}}\rho^2\bar{u}_t|u|^2udx=-\frac{1}{4}\frac{d}{dt}\int_{\mathcal{Q}}\rho^2|u|^4dx.
\end{eqnarray*}
On the other hand, due to the boundary conditions (\ref{eq2aux}):
\begin{eqnarray*}
-\frac{1}{2}\mathrm{Re}\int_{\mathcal{Q}}\rho^2\bar{u}_tu_{xx}dx&=&-\frac{1}{2}\mathrm{Re}\left[\rho^2u_x\bar{u}_t\right]_{-L}^{L}+\frac{1}{2}\mathrm{Re}\int_{\mathcal{Q}}\left(\rho^2\bar{u}_{t}\right)_xu_xdx\nonumber\\
&=&\frac{1}{2}\mathrm{Re}\int_{\mathcal{Q}}\rho^2\bar{u}_{tx}u_x+\mathrm{Re}\int_{\mathcal{Q}}\rho\rho'\bar{u}_tu_xdx\nonumber\\
&=&
\frac{1}{4}\frac{d}{dt}\int_{\mathcal{Q}}\rho^2|u_x|^2dx+\mathrm{Re}\int_{\mathcal{Q}}\rho\rho'\bar{u}_tu_xdx.
\end{eqnarray*}
Let us recall from \cite{BrezisFA, ZeiIIA}, that $H^2_0(\mathcal{Q})=\left\{u\in H^2(\mathcal{Q})\;:\;u=u_x=0\right\}$. 
Therefore, the boundary term $-\frac{1}{2}\mathrm{Re}\left[\rho^2u_x\bar{u}_t\right]_{-L}^{L}=0$, and we arrive to a weighted version of an energy balance equation:
\begin{eqnarray}
\label{eqn15}
\frac{1}{4}\frac{d}{dt}\left[\int_{\mathcal{Q}}\rho^2|u_x|^2dx-\int_{\mathcal{Q}}\rho^2|u|^4dx\right]=-\mathrm{Re}\int_{\mathcal{Q}}\rho\rho'\bar{u}_tu_xdx+\mathrm{Re}\int_{\mathcal{Q}}\mathrm{i}\gamma u\bar{u}_t\rho^2dx-\mathrm{Re}\int_{\mathcal{Q}}f\bar{u}_t\rho^2dx.
\end{eqnarray}
Moreover,  by solving  Eq. (\ref{eq1}) for $\bar{u}_t$, and then substituting into the first term of the right-hand side of (\ref{eqn15}), we get the following expression in terms of $u$ and its derivative $u_x$:
\begin{eqnarray}
\label{eqn16}
\mathrm{Re}\int_{\mathcal{Q}}\mathrm{i}\gamma u\bar{u}_t\rho^2dx&=&\mathrm{Re}\int_{\mathcal{Q}}\mathrm{i}\gamma u(-\frac{\mathrm{i}}{2}\bar{u}_{xx}-\mathrm{i}|u|^2\bar{u}-\gamma\bar{u}+\mathrm{i}\bar{f})\rho^2dx\nonumber\\
&=&\frac{1}{2}\mathrm{Re}\int_{\mathcal{Q}}\gamma u\bar{u}_{xx}\rho^2dx+\mathrm{Re}\int_{\mathcal{Q}}\gamma u|u|^2\bar{u}\rho^2dx-\mathrm{Re}\int_{\mathcal{Q}}\mathrm{i}\gamma^2u\bar{u}\rho^2dx-\mathrm{Re}\int_{\mathcal{Q}}\gamma u\bar{f}\rho^2dx\nonumber\\
&=&-\frac{\gamma}{2}\mathrm{Re}\int_{\mathcal{Q}}\bar{u}_x(\rho^2u)_xdx+\gamma\int_{\mathcal{Q}}\rho^2|u|^4dx-\gamma\mathrm{Re}\int_{\mathcal{Q}}\rho^2\bar{f}udx\nonumber\\
&=&-\frac{\gamma}{2}\int_{\mathcal{Q}}\rho^2|u_x|^2dx-\gamma\mathrm{Re}\int_{\mathcal{Q}}\rho\rho'u\bar{u}_xdx+\gamma\int_{\mathcal{Q}}\rho^2|u|^4dx-\gamma\mathrm{Re}\int_{\mathcal{Q}}\rho^2\bar{f}udx.
\end{eqnarray}
Similarly, for the third term of the right-hand side of (\ref{eqn15}), we have:
\begin{eqnarray}
\label{eqn17}
-\mathrm{Re}\int_{\mathcal{Q}}f\bar{u}_t\rho^2dx=\mathrm{Re}\int_{\mathcal{Q}}\rho^2f_t\bar{u}dx-\frac{d}{dt}\mathrm{Re}\int_{\mathcal{Q}}\rho^2f\bar{u}dx.
\end{eqnarray} 
Replacing both terms of the right-hand side of (\ref{eqn15}), by their equivalent expressions given in (\ref{eqn16}) and (\ref{eqn17}), we see that the former can be rewritten as  
\begin{eqnarray}
\label{eqn18}
&&\frac{d}{dt}\left[\frac{1}{4}\int_{\mathcal{Q}}\rho^2|u_x|^2dx-\frac{1}{4}\int_{\mathcal{Q}}\rho^2|u|^4dx+\mathrm{Re}\int_{\mathcal{Q}}\rho^2f\bar{u}dx\right]+\frac{\gamma}{2}\int_{\mathcal{Q}}\rho^2|u_x|^2dx-\gamma\int_{\mathcal{Q}}\rho^2|u|^4dx+\gamma\mathrm{Re}\int_{\mathcal{Q}}\rho^2\bar{f}udx\nonumber\\
&&=-\mathrm{Re}\int_{\mathcal{Q}}\rho\rho'\bar{u}_tu_xdx-\gamma\mathrm{Re}\int_{\mathcal{Q}}\rho\rho'u\bar{u}_xdx+\mathrm{Re}\int_{\mathcal{Q}}\rho^2f_t\bar{u}dx.
\end{eqnarray}
We define the functional $\mathcal{J}[u]$, as
$$
\mathcal{J}[u]=\frac{1}{4}\int_{\mathcal{Q}}\rho^2|u_x|^2dx-\frac{1}{4}\int_{\mathcal{Q}}\rho^2|u|^4dx+\mathrm{Re}\int_{\mathcal{Q}}\rho^2f\bar{u}dx.
$$
By adding, and simultaneously subtracting, in the left-hand side of (\ref{eqn18}), the terms $\frac{\gamma}{4}\int_{\mathcal{Q}}\rho^2|u_x|^2dx$ and $\frac{\gamma}{4}\int_{\mathcal{Q}}\rho^2|u|^4dx$, we get that $\mathcal{J}[u]$ satisfies the equation 
\begin{eqnarray}
\label{eqn19}
\frac{d}{dt}\mathcal{J}[u]+\gamma \mathcal{J}[u]+ \frac{\gamma}{4}\int_{\mathcal{Q}}\rho^2|u_x|^2dx=-\mathrm{Re}\int_{\mathcal{Q}}\rho\rho'\bar{u}_tu_xdx-\gamma\mathrm{Re}\int_{\mathcal{Q}}\rho\rho'u\bar{u}_xdx+\mathrm{Re}\int_{\mathcal{Q}}\rho^2f_t\bar{u}dx+\frac{3\gamma}{4}\int_{\mathcal{Q}}\rho^2|u|^4dx.
\end{eqnarray}
The integral terms of the right-hand side of (\ref{eqn19}), can be estimated as follows: we note first, that due to Theorem \ref{L1}, there exist constants $R'$ and $R''$, such that $\int_\mathcal{Q}|u_t|^2dx<R'$, and $||u||_{\infty}^2\leq R''$, respectively. Therefore, we have:
\begin{eqnarray*}
\left|\mathrm{Re}\int_\mathcal{Q}\rho\rho'\bar{u}_tu_xdx\right|	&\leq& \frac{\gamma}{16}\int_\mathcal{Q}\rho^2|u_x|^2dx+\frac{4\beta_2^2}{\gamma}\int_\mathcal{Q}|u_t|^2dx\leq \frac{\gamma}{16}\int_\mathcal{Q}\rho^2|u_x|^2dx+\frac{4\beta_2^2R'}{\gamma},\\
\gamma\left|\mathrm{Re}\int_\mathcal{Q}\rho\rho'u\bar{u}_xdx\right|	&\leq&
\frac{\gamma}{16}\int_\mathcal{Q}\rho^2|u_x|^2dx+4\beta_2^2\gamma\int_\mathcal{Q}|u|^2dx\leq \frac{\gamma}{16}\int_\mathcal{Q}\rho^2|u_x|^2dx+4\beta_2^2\gamma R,\\
\left|\mathrm{Re}\int_{\mathcal{Q}}\rho^2f_t\bar{u}dx\right|&\leq&\frac{1}{2}\int_{\mathcal{Q}}\rho^2|u|^2dx+\frac{1}{2}\int_{\mathcal{Q}}\rho^2|f_t|^2dx\leq\frac{R_0}{2}+\frac{1}{2}\int_{\mathcal{Q}}\rho^2|f_t|^2dx.\\
\frac{3\gamma}{4}\int_{\mathcal{Q}}\rho^2|u|^4dx&\leq&  \frac{3\gamma}{4}||u||_{\infty}^2\int_{\mathcal{Q}}\rho^2|u|^2dx\leq \frac{3\gamma R''R_0}{4}.
\end{eqnarray*}
Inserting all the above estimates into the right-hand side of (\ref{eqn19}), we get the differential inequality for $\mathcal{J}[u]$:
\begin{eqnarray}
\label{eqn20}
\frac{d}{dt}\mathcal{J}[u]+\gamma \mathcal{J}[u]+\frac{\gamma}{8}\int_{\mathcal{Q}}\rho^2|u_x|^2dx\leq  M_1,\;\;M_1=\frac{4\beta_2^2R'}{\gamma}+ 4\beta_2^2\gamma R+\frac{3\gamma R''R_0}{4}+\frac{R_0}{2}+\frac{1}{2}\int_{\mathcal{Q}}\rho^2|f_t|^2dx.
\end{eqnarray}
Omitting the term $\frac{\gamma}{8}\int_{\mathcal{Q}}\rho^2|u_x|^2dx\geq 0$ from the left-hand side of (\ref{eqn20}), and taking the supremum for $t\in [0,\infty)$ on the quantity $M_1$, we may derive the existence of a numerical constant $M_2>0$ (still depending on $\gamma$, on the norms of $u_0$ and $f_t$ in $L^{\infty}(\mathbb{R},L^2_{\rho^2}(\mathcal{Q}))$, but not on $t$ and $L$), such that 
$$
\mathcal{J}[u(t)]\leq \mathcal{J}[u_0]\exp(-\gamma t)+M_2(1-\exp(-\gamma t)).
$$
Evidently, the claimed estimate (\ref{eqn8}), holds with $R_1=R_0+M_2$.
\ \ \ $\Box$

With Theorem \ref{L1} and Lemmas \ref{LemW1} and \ref{LemW2} in hand, we consider the approximating initial-boundary value problem for the problem (\ref{eq1})-(\ref{eq2}) in the real line, as follows:
\begin{eqnarray}
\label{eq1L}
&&\mathrm{i}u_t^L+\frac{1}{2}u_{xx}^L+|u^L|^2u^L =f-\mathrm{i}\gamma u^L,
\;\;\forall\,x\in (-L,L),\;\;t>0,\\
\label{eq2L}
&& u^L(-L,t)=u^L(L,t)=0,\;\;\mbox{for all}\;\;t\geq 0,\\
\label{eq3L}
&& u^L(x,0)= u_0^L(x)\quad \forall\,x\in (-L,L).
\end{eqnarray}
Solutions of the approximating problem (\ref{eq1L})-(\ref{eq3L}), will define an approximating sequence,  converging to a weak solution of the problem (\ref{eq1})-(\ref{eq2}) as $L\rightarrow\infty$. Let us recall at this point, the weak formulation for the problem (\ref{eq1})-(\ref{eq2}).  Decomposing Eq.~(\ref{eq1}) in its real and imaginary parts, for $u=U+\mathrm{i}W$, and $f=g+\mathrm{i}h$, we consider weak solutions, at least in the class $u\in C([0,T], H^2(\mathbb{R}))$, $u_t\in C([0,T], L^2(\mathbb{R}))$, satisfying for the given initial condition $u_0\in H^2(\mathbb{R})\cap H^1_{\rho^2}(\mathbb{R})$, the following weak formulas:
\begin{eqnarray}
\label{weakU}
&&\int_{0}^T\int_{\mathbb{R}}W_t vdxdt-\frac{1}{2}\int_{0}^T\int_{\mathbb{R}}U_{xx}\,vdxdt-\int_{0}^T\int_{\mathbb{R}}U(U^2+W^2)vdxdt=-\int_{0}^T\int_{\mathbb{R}}gvdxdt-\gamma\int_{0}^T\int_{\mathbb{R}}Wvdxdt,\\
\label{weakW}
&&\int_{0}^T\int_{\mathbb{R}}U_t vdxdt+\frac{1}{2}\int_{0}^T\int_{\mathbb{R}}W_{xx}\,vdxdt+\int_{0}^T\int_{\mathbb{R}}W(U^2+W^2)vdxdt=\int_{0}^T\int_{\mathbb{R}}hvdxdt-\gamma\int_{0}^T\int_{\mathbb{R}}Wvdxdt,
\end{eqnarray} 
for all $v\in C_0^{\infty}([0,T]\times\mathbb{R})$.
Passing to the limit as $L\rightarrow\infty$, we shall use the following compactness Lemma.
\begin{lemma}
\label{LemW3}
The inclusion $H^2(\mathbb{R})\cap H^1_{\rho^2}(\mathbb{R})\subset H^1(\mathbb{R})$ is compact.
\end{lemma} \textbf{Proof:}	Let $\left\{u_n\right\}_{n=1}^{\infty}$ be a bounded sequence in $H^2(\mathbb{R})\cap H^1_{\rho^2}(\mathbb{R})$. Therefore, there exists a weakly convergent subsequence $u_{n_j}$, such that $u_{n_j}\rightharpoonup u$ in  $H^2(\mathbb{R})\cap H^1_{\rho^2}(\mathbb{R})$, as $j\rightarrow\infty$. Since the inclusion $H^2(\mathbb{R})\cap H^1_{\rho^2}(\mathbb{R})\subset H^1(\mathbb{R})$ is continuous, $u_{n_j}\rightharpoonup u$ also in $H^1(\mathbb{R})$. We will verify that $u_{n_j}\rightarrow u$, as   $j\rightarrow\infty$, i.e., strongly in $H^1(\mathbb{R})$. We observe that 
\begin{eqnarray}
\label{eq22a}
\|u_{n_j}-u\|^2_{H^1(\mathbb{R})}&=&\sum_{k=0}^1\int_{\mathbb{R}}|\partial^k(u_{n_j}-u)|^2dx=\sum_{k=0}^1\int_{\mathcal{Q}}|\partial^k(u_{n_j}-u)|^2dx+\sum_{k=0}^1\int_{\mathbb{R\setminus \mathcal{Q}}}|\partial^k(u_{n_j}-u)|^2dx\nonumber\\
&=&
\sum_{k=0}^1\int_{\mathcal{Q}}|\partial^k(u_{n_j}-u)|^2dx+\sum_{k=0}^1\int_{\mathbb{R\setminus \mathcal{Q}}}\rho^2\rho^{-2}|\partial^k(u_{n_j}-u)|^2dx\nonumber\\
&\leq&
\sum_{k=0}^1\int_{\mathcal{Q}}|\partial^k (u_{n_j}-u)|^2dx+\sup_{|x|>L}|\rho^{-2}(x)|\sum_{k=0}^1\int_{\mathbb{R\setminus \mathcal{Q}}}\rho^2|\partial^k(u_{n_j}-u)|^2dx\nonumber\\
&=&\sum_{k=0}^1\int_{\mathcal{Q}}|\partial^k(u_{n_j}-u)|^2dx+|\rho^{-2}(L)|\sum_{k=0}^1\int_{\mathbb{R\setminus \mathcal{Q}}}\rho^2|\partial^k (u_{n_j}-u)|^2dx.
\end{eqnarray}
Then, since $u_n-u\in H^2(\mathbb{R})\cap H^1_{\rho^2}(\mathbb{R})$, for all $n\in\mathbb{N}$, the inequality (\ref{eq22a}) implies that there exists a constant $C>0$, such that
$$
\|u_{n_j}-u\|^2_{H^1(\mathbb{R})}\leq \|u_{n_j}-u\|^2_{H^1(\mathcal{Q})}+\frac{C}{\rho^2(L)}.
$$
For arbitrary given $\epsilon>0$, we may choose $L$ sufficiently large, such that $\frac{C}{\rho^2(L)}<\epsilon/2$. On the other hand, the embedding  $H^2(\mathcal{Q})\subset H^1(\mathcal{Q})$ is compact. Hence,  $u_{n_j}\rightarrow u \in H^1(\mathcal{Q})$, as  $j\rightarrow\infty$, and we may also choose $j$ sufficiently large, such that $\|u_{n_j}-u\|^2_{H^1(\mathcal{Q})}<\epsilon/2$. Summarizing, for sufficiently large $j\in\mathbb{N}$ and $L\in\mathbb{R}$,  we have that $\|u_{n_j}-u\|^2_{H^1(\mathbb{R})}<\epsilon/2 +\epsilon/2=\epsilon$. Hence,  $u_{n_j}\rightarrow u$ in $H^1(\mathbb{R})$, and the claim of the lemma is proved. \ $\Box$

We may proceed to the statement and proof of the main result on the spatial decay estimates of the problem (\ref{eq1})-(\ref{eq2}), in the following
\begin{proposition}
\label{spatial}
Assume that the initial condition of the problem (\ref{eq1})-(\ref{eq2}), is such that $u_0\in  H^2_0(\mathbb{R})\cap H^1_{\rho^2}(\mathbb{R})$, and  that $f,\,f_t\in L^{\infty}(\mathbb{R},L^2_{\rho^2}(\mathbb{R}))$. Then, there exists a constant $K_1>0$, such that the density of the solution, decays at least, according the spatial decay estimate 
\begin{eqnarray}
\label{Sdec}
|u(x,t)|^2\leq \frac{K_1}{\rho^2(x)},\;\;\mbox{for all}\;\; t\geq 0.
\end{eqnarray} 
\end{proposition}
\textbf{Proof:} The initial condition $u_0\in  H^2_0(\mathbb{R})\cap H^1_{\rho^2}(\mathbb{R})$ can be approximated by an $L$-dependent sequence $u^{L}_0(x)$, when $L\rightarrow\infty$, defined as follows:  $u^{L}_0(x)=\Psi(|x|)u_0(x)$, where $\Psi(|x|)$ is a sufficiently smooth function with the cut-off properties
\begin{eqnarray*}
\Psi(|x|)=\left\{
\begin{array}{ll}
	1,\;\hbox{if $|x|\leq L$,}\\
	0,\;\hbox{if $|x|> L$.}
\end{array}
\right.
\end{eqnarray*}
Clearly,
\begin{eqnarray*}
\|u_0^{L}\|_{ H^2_0(\mathbb{R})}\leq \|u_0\|_{ H^2_0(\mathbb{R})},\;\;\;\;
\|u_0^{L}\|_{H^1_{\rho^2}(\mathbb{R})}\leq \|u_0\|_{H^1_{\rho^2}(\mathbb{R})}.
\end{eqnarray*}
Moreover, $u^{L}_0\rightarrow u_0$ in $H^1(\mathbb{R})$. Indeed,
\begin{eqnarray*}
\|u^L_0-u_0\|^2_{L^2}&=&\int_{|x|\leq L}\left|\Psi(|x|-1)\right|^2\;|u_0(x)|^2dx+\int_{|x|>L}\left|\Psi(|x|-1)\right|^2\;|u_0(x)|^2dx\\
&=&\int_{|x|>L}|u_0|^2dx\leq |\rho^{-2}(L)|\int_{|x|>L}\rho^2|u_0|^2dx\rightarrow 0\;\;\mbox{as}\;\;L\rightarrow\infty.
\end{eqnarray*}
Similarly, for the $L^2$-norm of the derivative we have:
\begin{eqnarray*}
	\|(u^{L}_{0})' -(u_{0})'\|^2_{L^2}&=&\int_{|x|\leq L}|\Psi(|x|)'u_0(x)+\Psi(|x|)u_0(x)'-u_0(x)'|^2dx\\&+&\int_{|x|> L }|\Psi(|x|)'u_0(x)+\Psi(|x|)u_0(x)'-u_0(x)'|^2dx \\
	&=&\int_{|x|>L}|\Psi(|x|)-1|^2|u_0(x)'|^2=\int_{|x|>L}|u_0(x)'|^2\\ 
	&\leq& |\rho^{-2}(L)|\int_{|x|>L}\rho^2(x)|u_0(x)'|^2dx\rightarrow 0\;\;\mbox{as}\;\;L\rightarrow\infty.
\end{eqnarray*}

Working as above, we may approximate higher order derivatives. Then, for $L\in\mathbb{N}$, we consider the approximating problem (\ref{eq1L})-(\ref{eq3L}), with the initial condition $u^L(x,0)= u_0^L(x)$, for $t\in [0, T]$, for arbitrary $T>0$. There exists a unique solution of the above approximating problem satisfying the estimate described in Theorem \ref{L1}, and the weighted estimates described in Lemmas \ref{LemW1} and \ref{LemW2}. Since the latter are uniform on $L$, we may extend the solution $u^L$ for $x\in \mathbb{R}$, as 
$$
\tilde{u}^L(x,t)=\left\{
\begin{array}{ll}
u^L(x,t),\;\hbox{if $|x|\leq L$,}\\
0,\;\;\;\;\;\;\;\;\;\;\,\,\hbox{if $|x|> L$.}
\end{array}
\right.
$$ 
The extended sequence $\tilde{u}^L$, is bounded in $C([0,T], H^2(\mathbb{R})])$, with $\tilde{u}_t^L$ bounded in $C([0,T], L^2(\mathbb{R})])$. Combining Theorem \ref{L1}, and Lemma \ref{LemW2}, we derive that $\tilde{u}^L$ is bounded in $L^{\infty}([0,T], H^2(\mathbb{R})\cap H^1_{\rho^2}(\mathbb{R}))$. It follows that there exists a subsequence $u^{L_j}$ such that
\begin{eqnarray}
\label{eqnn25}
&& u^{L_j}\stackrel{\ast}{\rightharpoonup}u^{\infty},\;\; \hbox{in}\;\;L^{\infty}([0,T], H^2(\mathbb{R})),\\
\label{eq26}
&& u_t^{L_j}\stackrel{\ast}{\rightharpoonup}u_t^{\infty},\;\; \hbox{in}\;\;L^{\infty}([0,T], L^2(\mathbb{R})).
\end{eqnarray}
Moreover, due to the compactness Lemma \ref{LemW3}, if applying Simon's compactness results \cite[Corollary 4,
p.85]{sim90}, on the inclusions $H^2(\mathbb{R})\cap H^1_{\rho^2}(\mathbb{R})\subset H^1(\mathbb{R})\subset L^2(\mathbb{R})$, we derive that
\begin{eqnarray}
\label{eq27}
u^{L_j}\rightarrow u^{\infty},\;\; \hbox{in}\;\;C([0,T], H^1(\mathbb{R})).
\end{eqnarray}
The convergence relations (\ref{eqnn25}) and (\ref{eq27}), suffice in order to pass to the limit to the nonlinear terms of the weak formulas (\ref{weakU}) and (\ref{weakW}), due to the continuity of  $|u|^2=U^2+W^2$. For instance, we have
\begin{eqnarray}
\label{eq28}
W^{L_j}|u^{L_j}|^2\rightharpoonup W^{\infty}|u^{\infty}|^2,\;\;U^{L_j}|u^{L_j}|^2\rightharpoonup U^{\infty}|u^{\infty}|^2,\;\; \hbox{in}\;\;L^2([0,T], H^1(\mathbb{R})).
\end{eqnarray}
With the convergence relations (\ref{eqnn25})-(\ref{eq28}) in hand, and by applying the restriction procedure of \cite[Theorem 1.3, pg. 228]{Bab90} or \cite[Proposition 3.2, pg. 192]{NK1999}, we may conclude that $u(x,t)=u^{\infty}(x,t)$ is a weak solution of the problem (\ref{eq1})-(\ref{eq2}),
still satisfying the same estimates (\ref{eq4}), (\ref{eqn14}), and (\ref{eqn14es}).  Furthermore, these estimates, and the boundedness of $|\rho'(x)|$ from (\ref{eqn5}), imply the existence of some constants $c_1,\, c_2,\, R_2>0$ such that
$$
\int_{\mathbb{R}}|\partial_x(\rho u)|^2dx=\int_{\mathbb{R}}|\partial_x(\rho u)|^2dx=
\int_{\mathbb{R}}|\rho' u+\rho u_x|^2dx\leq c_1\int_{\mathbb{R}}|u|^2dx+c_2\int_{\mathbb{R}}\rho^2|u_x|^2dx\leq R_2,
$$
where the constant $R_2$ depends only on $\beta_2,\, R,\, R_0$, and $R_1$. To derive the density estimate (\ref{Sdec}), we apply the Agmon's interpolation inequality  
\begin{eqnarray}
\label{eqAg}
\|\psi\|_{L^{\infty}}^2\leq \|\psi\|_{L^2}\|\psi_x\|_{L^2},\;\;\mbox{for all}\;\;\psi\in C_0^{\infty}(\mathbb{R}),
\end{eqnarray}
on the function $\psi(x,t)=\rho(x)u(x,t)$: we have
\begin{eqnarray*}
\|\rho  u\|^2_{\infty}\leq \|\rho u\|_{L^2}\|\partial_x(\rho u)\|_{L^2}	\leq K_1,
\end{eqnarray*}
with $K_1=\sqrt{R_0R_2}$, from which the density estimate (\ref{Sdec}), readily follows. \ $\Box$
\subsection{Time-decay estimates}
\label{SUSB}
To derive estimates on the rate of decay of  the solutions of the problem (\ref{eq1})-(\ref{eq2}) with respect to time, we apply the time-weighted energy method proposed in \cite{HS}. This method was originally introduced for the proof of time-decay  estimates for the solutions of the Navier-Stokes equations. 
We will consider weight functions $\phi(t)$ with the following properties:
 \begin{eqnarray}
 \label{wc}
 \phi(t)>0,\;\;\forall\; t\in\mathbb{R},\;\;\lim_{|t|\rightarrow\infty}\phi(t)=+\infty,&& 0<\delta_1<|\dot{\phi}(t)|<\delta_2,\;\;\forall\;t\in\mathbb{R},\;\;\delta_1,\delta_2=\mathrm{const.},\\
\label{wcalt}
\hbox{or alternatively}, && 0<\delta_1'<|\dot{\phi}(t)|<\delta_2'|\phi(t)|,\;\;\forall\;t\in\mathbb{R},\;\;\delta_1',\delta_2'=\mathrm{const.}
 \end{eqnarray}
 As it will be proved in the following proposition, either cases (\ref{wc}) or (\ref{wcalt}), distinguish between an unconditional or a conditional decaying behavior of the solutions, with respect to the damping strength $\gamma>0$.
\begin{proposition}
\label{P1}
Let $u_0\in H^2_0(\mathcal{Q})$,  where either  $\mathcal{Q}=\mathbb{R}$ and the solution $u$ of (\ref{eq1}), satisfies the vanishing boundary conditions (\ref{eq2}), or $\mathcal{Q}=(-L,L)$, and the solution $u$ of (\ref{eq1}), satisfies the Dirichlet boundary conditions (\ref{eq2aux}). We assume that the forcing term is such that $f,\,f_t\in L_{\mathrm{loc}}^{\infty}(\mathbb{R},L^2(\mathcal{Q}))$, and in addition,  that
\begin{eqnarray}
\label{fd}
\int_0^{\infty}\int_{\mathcal{Q}}\phi^2(t)|f(x,t)|^2dxdt<\infty.
\end{eqnarray}
\begin{enumerate}
	\item
(Unconditional time-decay with respect to $\gamma$). Consider a weight function $\phi(t)$, satisfying condition (\ref{wc}). 	Then, there exists a constant $K_2>0$ independent of $t$, such that the density of the solution of the problem (\ref{eq1})-(\ref{eq2}), decays in time at least, according to the estimate
\begin{eqnarray}
\label{test}
|u(x,t)|^2\leq\frac{K_2}{\phi(t)}\;\;\mbox{for all}\;\;x\in\mathcal{Q},\;\;t>0,\;\;\hbox{for all}\;\;\gamma>0.
\end{eqnarray} 
\item
(Conditional time-decay with respect to $\gamma$). Assume alternatively that the weight function, $\phi(t)$ satisfies the condition (\ref{wcalt}),  Then the density time-decaying estimate (\ref{test}) is still valid, for
$\gamma\geq\delta_2'$.
\end{enumerate}
\end{proposition}
\textbf{Proof}:  We apply the method of \cite[Sec. 3.4, pg. 334]{HS}: Multiplying the equation (\ref{eq1}) and the boundary conditions (\ref{eq2}), by $\phi(t)$, and noticing that $\phi(0)u(x,0)=\phi(0)u_0$, we observe that the function $\phi(t)u(x,t)$ satisfies the weighted equation
\begin{eqnarray}
\label{eq17}
\mathrm{i}{{(u\phi)}_{t}}+\frac{1}{2} {{u}_{xx}}\phi+|u|^2u\phi =f-\mathrm{i}\gamma u\phi+\mathrm{i}u\dot{\phi}.
\end{eqnarray}
In the case where $\mathcal{Q}=\mathbb{R}$,  the function $\phi u$ satisfies the boundary conditions
$$	
\lim_{|x|\rightarrow\pm \infty}\phi(t) u(x,t)=0,\;\;\mbox{for all}\;\;t\geq 0,
$$
and the initial condition
$$
\phi (0) u (x,0)=\phi(0) u_0(x),\quad \forall\,x\in\mathcal{Q}.
$$
We work as in Lemma \ref{L1}, i.e., multiplying Eq. (\ref{eq17}) by $\phi\bar{u}$ and integrating over $\mathcal{Q}$, keeping the imaginary parts of the resulting equation. Then $\phi u$ satisfies the weighted balance law
\begin{eqnarray}
\label{eq20}
\frac{1}{2}\frac{d}{dt}\int_{\mathcal{Q}}\phi^2(t)|u|^2dx+\gamma\int_{\mathcal{Q}}\phi^2(t)|u|^2dx=\mathrm{Im}\int_{\mathcal{Q}}\phi^2(t)\,f\bar{u}dx+\int_{\mathcal{Q}}\phi(t)\dot{\phi}(t)|u|^2dx.
\end{eqnarray}
Integrating Eq. (\ref{eq20})  in the interval $[0,t]$, for arbitrary $t>0$, we obtain the equation
\begin{eqnarray}
\label{eq21}
\frac{1}{2}\phi^2(t)\int_{\mathcal{Q}}|u|^2dx+\gamma\int_{0}^{t}\int_{\mathcal{Q}}\phi^2|u|^2dxd\tau=\frac{1}{2}\phi^2(0)\int_{\mathcal{Q}}|u_0|^2dx+\mathrm{Im}\int_0^t\int_{\mathcal{Q}}\phi^2\,f\bar{u}dxd\tau+\int_0^t\int_{\mathcal{Q}}\phi\dot{\phi}|u|^2dxd\tau
\end{eqnarray}
Handling of the last term of equation (\ref{eq21}), shall distinguish between the two cases stated above.
\begin{enumerate}
\item \textit{(Unconditional time-decay with respect to $\gamma$)}: Using the assumption (\ref{wc}), on the weight function $\phi(t)$, the last term of (\ref{eq21}) is estimated as
\begin{eqnarray*}
	\left|\int_0^t\int_{\mathcal{Q}}\phi\dot{\phi}|u|^2dxd\tau\right|\leq \delta_2\int_0^t\int_{\mathcal{Q}}\phi|u|^2dxd\tau\leq \frac{\gamma}{4}\int_{0}^{t}\int_{\mathcal{Q}}\phi^2|u|^2dxd\tau+\frac{\delta_2^2}{\gamma}\int_{0}^{t}\int_{\mathcal{Q}}|u|^2dxd\tau,
\end{eqnarray*}
while for its second term
\begin{eqnarray}
\label{eq212nd}
	\left|\int_{0}^{t}\int_{\mathcal{Q}}\phi^2 f\bar{u}dxd\tau\right|
	\leq\frac{1}{\gamma}\int_{0}^{t}\int_{\mathcal{Q}}\phi^2|f|^2dxd\tau
	+\frac{\gamma}{4}\int_{0}^{t}\int_{\mathcal{Q}}\phi^2|u|^2dxd\tau.
\end{eqnarray}
Therefore, Eq. (\ref{eq21}) becomes the inequality
\begin{eqnarray}
\label{eq22}
\frac{1}{2}\phi^2(t)\int_{\mathcal{Q}}|u|^2dx+\frac{\gamma}{2}\int_{0}^{t}\int_{\mathcal{Q}}\phi^2|u|^2dxd\tau\leq&& \frac{1}{2}\phi^2(0)\int_{\mathcal{Q}}|u_0|^2dx\nonumber\\
&&+\frac{1}{\gamma}\int_{0}^{t}\int_{\mathcal{Q}}\phi^2|f|^2dxd\tau+\frac{\delta_2^2}{\gamma}\int_{0}^{t}\int_{\mathcal{Q}}|u|^2dxd\tau.
\end{eqnarray}
Since by Theorem \ref{L1}, the solution $u$ is in the class $C([0,\infty) H^2(\mathbb{R})])$,  and the assumption (\ref{fd}) also holds,  both allow us to define the constant $K_0$, as
\begin{eqnarray*}
	K_0=\frac{1}{2}\phi^2(0)\int_{\mathcal{Q}}|u_0|^2dx+\frac{1}{\gamma}\int_{0}^{\infty}\int_{\mathcal{Q}}\phi^2|f|^2dxdt+\frac{\delta_2^2}{\gamma}\int_{0}^{\infty}\int_{\mathcal{Q}}|u|^2dxdt.
\end{eqnarray*}
Then, inequality (\ref{eq22}), directly implies the decay of the $L^2$-norm of the solution, according to 
\begin{eqnarray}
\label{eq23}
||u(x,t)||_2^2\leq \frac{2K_0}{\phi^2(t)}.
\end{eqnarray}
Eventually, the uniform bound (\ref{eq4}) and the decay estimate (\ref{eq23}), if inserted in Agmon's inequality (\ref{eqAg}),  imply that
\begin{eqnarray*}
	|u(x,t)|^2\leq \|u(t)\|_{\infty}^2\leq \frac{K_2}{\phi(t)},\;\;\mbox{for all}\;\;x\in\mathcal{Q},\;\;t>0.
\end{eqnarray*}
i.e., the decay estimate (\ref{test}), with the constant $K_2$ depending on $K_0$ and $R_0$.
\item \textit{(Conditional time-decay with respect to $\gamma$)}: Using the assumption (\ref{wcalt}), on the weight function $\phi(t)$, the last term of (\ref{eq21}) is estimated as
\begin{eqnarray*}
	\left|\int_0^t\int_{\mathcal{Q}}\phi\dot{\phi}|u|^2dxd\tau\right|\leq \delta_2'\int_0^t\int_{\mathcal{Q}}\phi^2|u|^2dxd\tau,
\end{eqnarray*}
while for its second term, we keep the estimate (\ref{eq212nd}). Therefore, in the case of a weight function $\phi(t)$ satisfying (\ref{wcalt}), we see that  Eq. (\ref{eq21}) becomes the inequality
\begin{eqnarray}
\label{eq22alt}
\frac{1}{2}\phi^2(t)\int_{\mathcal{Q}}|u|^2dx+\left(\gamma-\delta_2'\right)\int_{0}^{t}\int_{\mathcal{Q}}\phi^2|u|^2dxd\tau\leq \frac{1}{2}\phi^2(0)\int_{\mathcal{Q}}|u_0|^2dx+\frac{1}{2\gamma}\int_{0}^{t}\int_{\mathcal{Q}}\phi^2|f|^2dxd\tau.
\end{eqnarray}
Hence, in this case, for $\gamma\ge\delta'_2$, the assumption (\ref{fd}) on the driver  allows us to define from \eqref{eq22alt}, the constant:
\begin{eqnarray*}
	K_0=\frac{1}{2}\phi^2(0)\int_{\mathcal{Q}}|u_0|^2dx+\frac{1}{2\gamma}\int_{0}^{\infty}\int_{\mathcal{Q}}\phi^2|f|^2dxdt.	
\end{eqnarray*}
For the rest steps of the proof we proceed exactly as in the case 1.\ \ \ $\Box$
\end{enumerate}
\subsection{Spatiotemporal decaying estimates}
Clearly, the derivation of generic spatiotemporal decaying estimates, simply comes by the suitable combination of the conditions described in Propositions \ref{spatial} and \ref{P1}.  Such a combination is summarized in
\begin{theorem}
	\label{spti}
Let $u_0\in  H^2_0(\mathbb{R})\cap H^1_{\rho^2}(\mathbb{R})$, and  assume that $f,\,f_t\in L^{\infty}(\mathbb{R}^+,L^2_{\rho^2}(\mathbb{R}))$ satisfies the condition (\ref{fd}). Then, the density of the solution of the problem (\ref{eq1})-(\ref{eq2}), spatiotemporally decays, at least, according to the rates described by the estimates  stated in Propositions \ref{spatial} and \ref{P1}, e.g.  the spatial decaying estimate (\ref{Sdec}) and the time-decaying estimate (\ref{test}), are simultaneously valid.	
\end{theorem}
\paragraph*{\textbf{Formal examples of drivers/weights and the relevant spatiotemporal decaying estimates.}}
We conclude this section, by discussing some  formal examples of the weight functions  satisfying the conditions stated in Theorem \ref{spti}, and the relevant induced estimates.  

The first example for the non-autonomous driver possesses the form of a Gaussian function centered at $(0, 0)$ and  having spreads $\sigma_x,\;\sigma_t>0$, with respect to the space and time variable, respectively \cite{LetArx1}:
\begin{eqnarray}
	\label{wlf}
	f(x,t)&=&g(x,t)\exp(\mathrm{i}\Theta),\;\;\hbox{where}\;\;g(x,t)=\sqrt{2}\,\Gamma\exp\left(-\frac{x^2}{2\sigma_x^2}-\frac{t^2}{2\sigma_t^2}\right).
\end{eqnarray}
The function (\ref{wlf}), serves as simple phenomenological example, of a  spatio-temporally, exponentially localized forcing, of amplitude  $\Gamma>0$. In our study we will consider only the  case of $\Theta=\frac{\pi}{4}$. It should be warned that the phase factor $\Theta$ may have an important role in the dynamics of the dissipative system, and it will be discussed elsewhere.

The second example for the driver, will have the form: 
\begin{eqnarray}
\label{wlf1}
F(x,t)&=&G(x,t)\exp(\mathrm{i}\Theta),\;\;\hbox{where}\;\;G(x,t)=\frac{\sqrt{2}\, \Gamma}{X(x;\delta_x)T(t;\delta_t)},\;x\in\mathbb{R},\;\;t>0,\nonumber\\
X(x;\delta_x)&=&\left[1+\left|\frac{x}{\delta_x}\right|+\frac{\theta x^2}{\delta_x^2}\right]^2,\;\;T(x;\delta_t)=\left(1+\frac{ t}{\delta_t}+\frac{\omega t^2}{\delta_t^2}\right)^2\;\;t\geq 0.
\end{eqnarray}
This time, the function (\ref{wlf1}) serves as an example of a spatio-temporally, weakly localized forcing, decaying at an algebraic rate.  Its rate of decay, both in space and time, is at most quartic, as determined by the constants $\theta,\, \omega\geq 0$.  The constants $\delta_x,\delta_t$ measure the localization of the forcing with respect to the space and time variable, respectively.   

The simplest example of a spatial-weight which satisfies condition (\ref{eqn5}),  is given by the function: 
\begin{eqnarray}
\label{eqn6}
\rho(x)=1+\left|\frac{x}{x_0}\right|.
\end{eqnarray}
For the weight (\ref{eqn6}), the condition (\ref{eqn5}) is satisfied for $\beta_2=\frac{1}{x_0}$. 

A natural example for a time-weight $\phi(t)$  can be defined, by taking into account, that the time-decaying rate should depend on the damping parameter $\gamma>0$; this is justified in the unforced case, $f\equiv 0$, where  solutions decay exponentially with respect of $\gamma$. Asking for a potential algebraic decay in time, we may consider the simplest example of weight functions, satisfying  condition (\ref{wc}), or  its alternative (\ref{wcalt}): 
\begin{eqnarray}
\label{wct}
\phi(t)=\left(1+\frac{\gamma t}{t_0}\right)^{\kappa},\;\;t_0>0,\;\;\kappa\geq 1,\;\;\kappa=\mathrm{const.}
\end{eqnarray}
Evidently, the case of condition (\ref{wc}) corresponds to $\kappa=1$, with the constant  $\delta_2=\frac{\gamma}{t_0}$,  while the case of its alternative (\ref{wcalt}), corresponds to $\kappa>1$, with the constant $\delta_2'=\frac{\gamma\kappa}{t_0}$. In the latter case, it is interesting to observe  that the conditional decay restriction is actually implemented  not on the damping strength $\gamma$, but on the decay rate $\kappa$, and the parameter $t_0$: as suggested from the second term on the left-hand side of inequality (\ref{eq22alt}), the requirement $$\gamma-\delta_2'=\gamma\left(1-\frac{\kappa}{t_0}\right)\geq 0,$$
applies for $t_0\geq \kappa$. 

It is also important to discuss the conditions of Theorem \ref{spti}.  It is evident, that both the forcing (\ref{wlf}) and the weight (\ref{wct}) will satisfy the  integral condition (\ref{fd}),  due to the exponential localization of $f$.  Then, it follows from the decaying estimate (\ref{test}),  that the density may decay  in time, at least with an algebraic rate $\kappa>1$,
\begin{eqnarray}
\label{eq25c}
|u(x,t)|^2\leq\frac{K_2}{\left(1+\frac{\gamma t}{t_0}\right)^{\kappa}},\;\;x\in\mathbb{R},\;\;\hbox{for all $\gamma>0$,\;\; with}\;\;\;\; t_0\geq\kappa.
\end{eqnarray}
Also, it  straightforward to see that $f,f_t\in L^{\infty}(\mathbb{R}^+,L^2_{\rho^2}(\mathcal{Q}))$, ensuring the spatial decaying estimate  (\ref{Sdec}) of Theorem \ref{spti}. On the other hand, concerning the weakly localized forcing $F$,  when $\omega=0$ and $\theta=0$, it decays at least quadratically in space and time, and obviously, the condition (\ref{fd}) is satisfied for the weight function (\ref{wct}) with $\kappa=1$. 
Then, it follows from the decaying estimate (\ref{test}),  that the density decays in time, at least with a linear rate 
$$
|u(x,t)|^2\leq\frac{K_2}{\left(1+\gamma t\right)},\;\;x\in\mathbb{R}\;\;\hbox{for all}\;\;\;\;\gamma>0.
$$
For an at least, quadratic decay rate $\kappa\geq 2$ in the estimate (\ref{eq25c}), formally one has to ensure condition (\ref{fd}).  We  may set $\omega>0$ in $f(x,t)$, so that its time-decaying rate will be quartic. Furthermore,  in this case, clearly  $F,F_t\in L^{\infty}(\mathbb{R}^+,L^2_{\rho^2}(\mathcal{Q}))$, as needed for the validity of the spatial decay estimate (\ref{Sdec}) stated in  Theorem \ref{spti}.

The relevance of Theorem \ref{spti} and of the above examples, with the dynamics of the solutions of the problem (\ref{eq1})-(\ref{eq2}), will be explored by suitable numerical experiments, whose results are reported in the next section.
\section{Numerical investigations}\label{SIII}

In this section, we  perform  numerical simulations, aiming to investigate  the potential relevance and impact of the generic  spatio-temporal decaying estimates of Theorem \ref{spti}, in analyzing the dynamics of the problem (\ref{eq1})-(\ref{eq2}). As in \cite{LetArx1,LetArx2}, we will investigate the dynamics for vanishing initial conditions of the following forms: the algebraically (at a quadratic rate) decaying
\begin{eqnarray}
	\label{ic1}
	u_0(x)=\frac{1}{1+x^2},
\end{eqnarray}
and the exponentially decaying one %
\begin{eqnarray}
\label{ic2exp}
u_0(x)=\mathrm{sech}(x).
\end{eqnarray}
The initial condition (\ref{ic2exp}) resembles the profile of a bright soliton. Both initial conditions (\ref{ic1}) and (\ref{ic2exp}), are  satisfying the assumption of Theorem \ref{spti}, $u_0\in H^2_0(\mathbb{R})\cap H^1_{\rho^2}(\mathbb{R})$. 


For these numerical investigations, we have considered a Chebyshev polynomial pseudospectral scheme \cite{Trefethen} for the spatial integration, while, for the integration with respect to time we have used a 4th –- 5th order adaptive-step Runge-Kutta method (see e.g.~\cite{numrec}). 

Note that, in our numerical study, we have considered a system of finite spatial length $[-L,L]$ with $x(-L)=x(L)=0$. This fact imposes a numerical error, since, for both the initial conditions \eqref{ic1} and \eqref{ic2exp}, this condition is only satisfied asymptotically. Nevertheless, since we have considered $L\geqslant400$, in the case of initial condition \eqref{ic1} the initial error is of $\mathcal{O}(10^{-6})$, which poses no observable effects for the time intervals considered. On the other hand, for the condition \eqref{ic2exp}, the initial error is by far smaller than the accuracy used in the calculations.

\subsection{Gaussian driver}
\begin{figure}[tbh!]
	\hspace{-0.5cm}\includegraphics[scale=0.18]{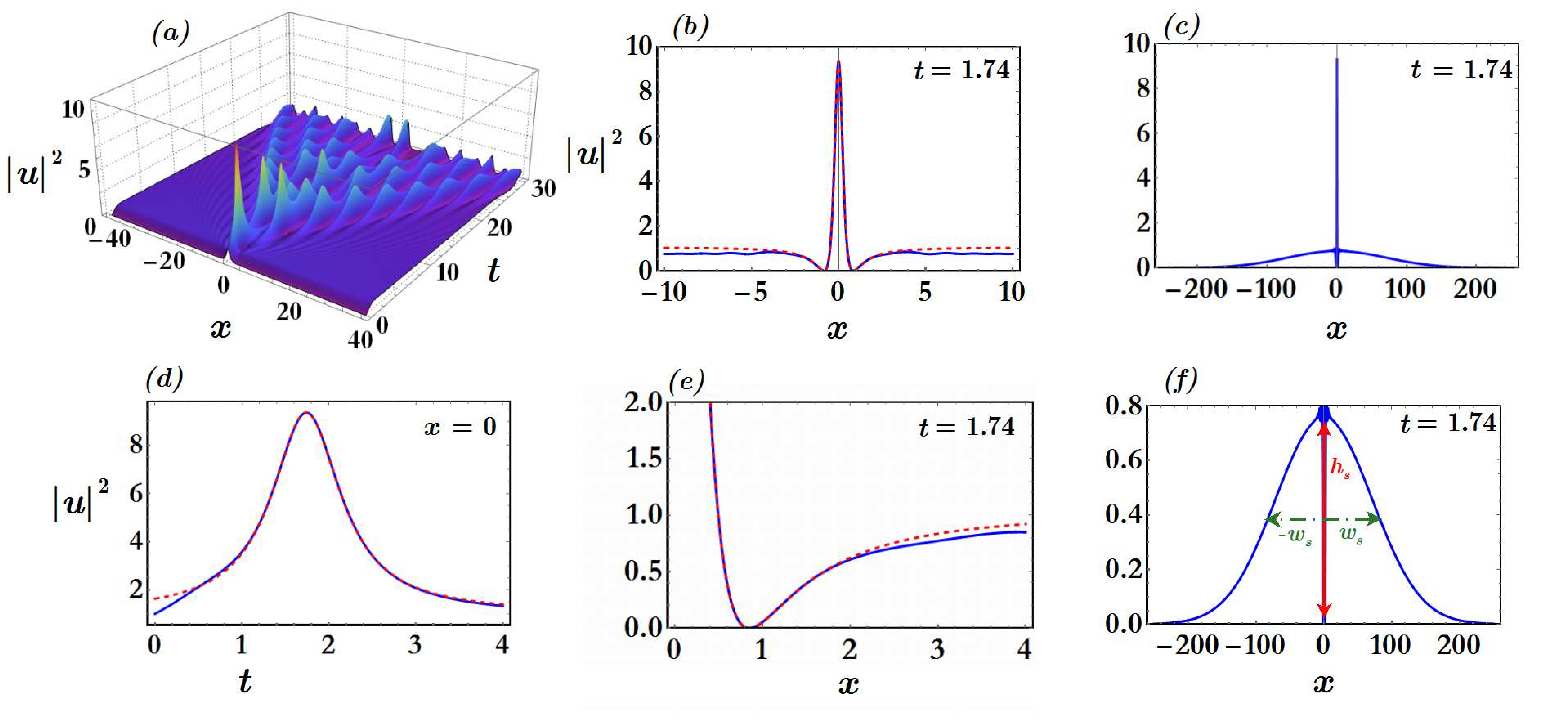}
	\caption{(Color Online) Dynamics of the algebraically decaying initial condition \eqref{ic1}, in the presence of the Gaussian driver \eqref{wlf}. Parameters $\gamma=0.01$, $\Gamma=1$, $\sigma_x=100$, $\sigma_t=0.5$, $L=500$. Top row: Panel (a) depicts a 3D-graph of the spatiotemporal evolution of the density $|u(x,t)|^2$, for $t\in [0,30]$.  Panel (b) shows a profile of the density at $t=1.74$ [solid (blue) curve], against the corresponding profile $u_{\mbox{\tiny PS}}(x,1.74;1.74;1.07)$, of the PRW  (\ref{PRW}) [dashed (red) curve]. Panel (c) portrays an expanded view of the snapshot shown in panel (b). Bottom row:  Panel (d) shows the evolution of the density of the center, $|u(0,t)|^2$,  against the evolution of the density of the center of  $u_{\mbox{\tiny PS}}(x,t;1.74;1.07)$. Panel  (e): a detail of the spatial profile of the numerical PRW-type event at $t^*=1.74$, close to the right of the two symmetric minima of the profile $u_{\mbox{\tiny PS}}(x,1.74;1.74;1.07)$ of the exact PRW  (\ref{PRW}). Panel (f) shows the profile of the emerged decaying support at $t=1.74$.}
	\label{figure1}
\end{figure}

We start our presentation with the case of the exponentially localized forcing. For completeness, we recall some of the numerical results presented in \cite{LetArx1}.  They are summarized in Fig.~\ref{figure1}, discussing the dynamics of the initial condition \eqref{ic1}.  
Panel (a) of the top row shows a 3D-graph of the spatiotemporal evolution of the density $|u(x,t)|^2$, for $t\in[0,30]$. The dynamics are found to share characteristics of the semi-classical limit of the integrable NLS, however with some important differences, as analyzed in \cite{LetArx1}. The first peak emerged in panel (a) corresponds to a rogue wave, strongly reminiscent of a PRW.  This fact is clarified in panel (b): The profile of the numerical solution at time $t=1.74$ [solid (blue) curve], is compared against the profile $u_{\mbox{\tiny PS}}(x,1.74;1.74;1.07)$ {\em of the analytical PRW-solution of the integrable NLS} [dashed (red) curve]: 
\begin{eqnarray}
	\label{PRW}
u_{\mbox{\tiny
			PS}}(x,t;t_0;P_0):=\sqrt{P_0}\left\{1-\frac{4\left[1+2\mathrm{i}P_0(t-t_0)\right]}{1+4P_0x^2+4P_0^2(t-t_0)^2}\right\}\mathrm{e}^{\mathrm{i}P_0(t-t_0)},
\end{eqnarray}
Note that the solution \eqref{PRW} is time translated at $t=t_0$ (here $t_0=1.74$) and is supported on a finite background of amplitude $P_0=1.07$.  Panel (c) shows an expanded view of the profile of the numerical solution at $t=1.74$, revealing the remarkable feature, that the PRW-type waveform is sustained on the top of a decaying support. Panel (d) of the bottom row is comparing the evolution of the density of the center $|u(0,t)|^2$, against the corresponding one of the PRW-soliton $u_{\mbox{\tiny PS}}(x,1.74;1.74;1.07)$. The half-length interval is increased to $L=500$ (if compared with the study of \cite{LetArx1}, where $L=250$); the time-growth and time decay of $|u(0,t)|^2$, is almost indistinguishable from the corresponding one of the PRW. The spatial decay of the PRW is also closely preserved, up to $|x|\sim 2$ far from the symmetric minima, as shown in panel (e). Panel (f) is magnifying the profile of the decaying support at $t=1.74$. The amplitude of the support is denoted by $h_s$,  while the half-width at half-maximum of the support 
is denoted by $w_s$. 
  
\begin{figure}[tbh!]
	\hspace{-0.5cm}\includegraphics[scale=0.18]{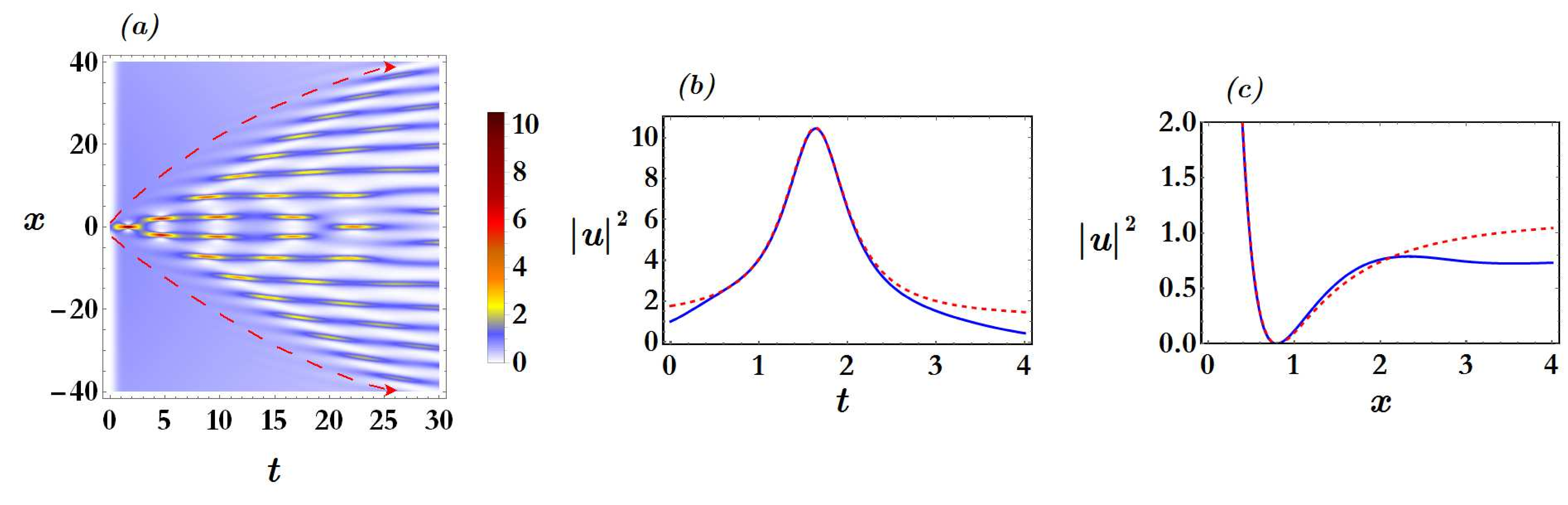}
	\caption{(Color Online) Dynamics of the exponentially decaying initial condition \eqref{ic2exp}, in the presence of the Gaussian driver \eqref{wlf}. Parameters $\gamma=0.01$, $\Gamma=1$, $\sigma_x=100$, $\sigma_t=0.5$, $L=500$. Panel (a) shows a contour plot of the spatiotemporal evolution of the density. Panel (b) shows the evolution of the density of the center, $|u(0,t)|^2$,  against the evolution of the density of the center of  $u_{\mbox{\tiny PS}}(x,t;1.63;1.16)$ . Panel  (c): a detail of the spatial profile of the numerical PRW-type event at $t^*=1.63$, close to the right of the two symmetric minima of the profile $u_{\mbox{\tiny PS}}(x,1.63;1.63;1.16)$ of the exact PRW  (\ref{PRW}).}
	\label{figure2}
\end{figure}

The decaying rate of the initial condition affects the proximity of the first emerged PRW-type event to the analytical PRW-solution. This is verified by the results of a new study on the dynamics of the exponentially decaying initial condition \eqref{ic2exp}. The dynamics, in this case,  is very similar to the one observed in Fig.~\ref{figure1}, as depicted in the contour plot of the spatio-temporal evolution of the density illustrated in panel (a) of Fig.~\ref{figure2}.  However, the numerical PRW-type event has a growing and decay rate which remains proximal to the analytical one for a smaller time interval, as shown in panel  (b) of Fig.~\ref{figure2}. The same holds for the spatial profile of the numerical event, as shown in panel (c), showing that the proximity persists mainly to the core of the analytical PRW. These differences can be understood by the fact that the algebraic initial condition \eqref{ic1} shares the same quadratic decaying rate of the analytical PRW. Thus, it is expected that the numerical solution shall preserve the space-time localization rates for larger time intervals, than the one initiated from the $\mathrm{sech}$-profiled initial datum which possesses an exponential decay rate. 
\begin{figure}[tbh!]
	\hspace{-0.5cm}\includegraphics[scale=0.18]{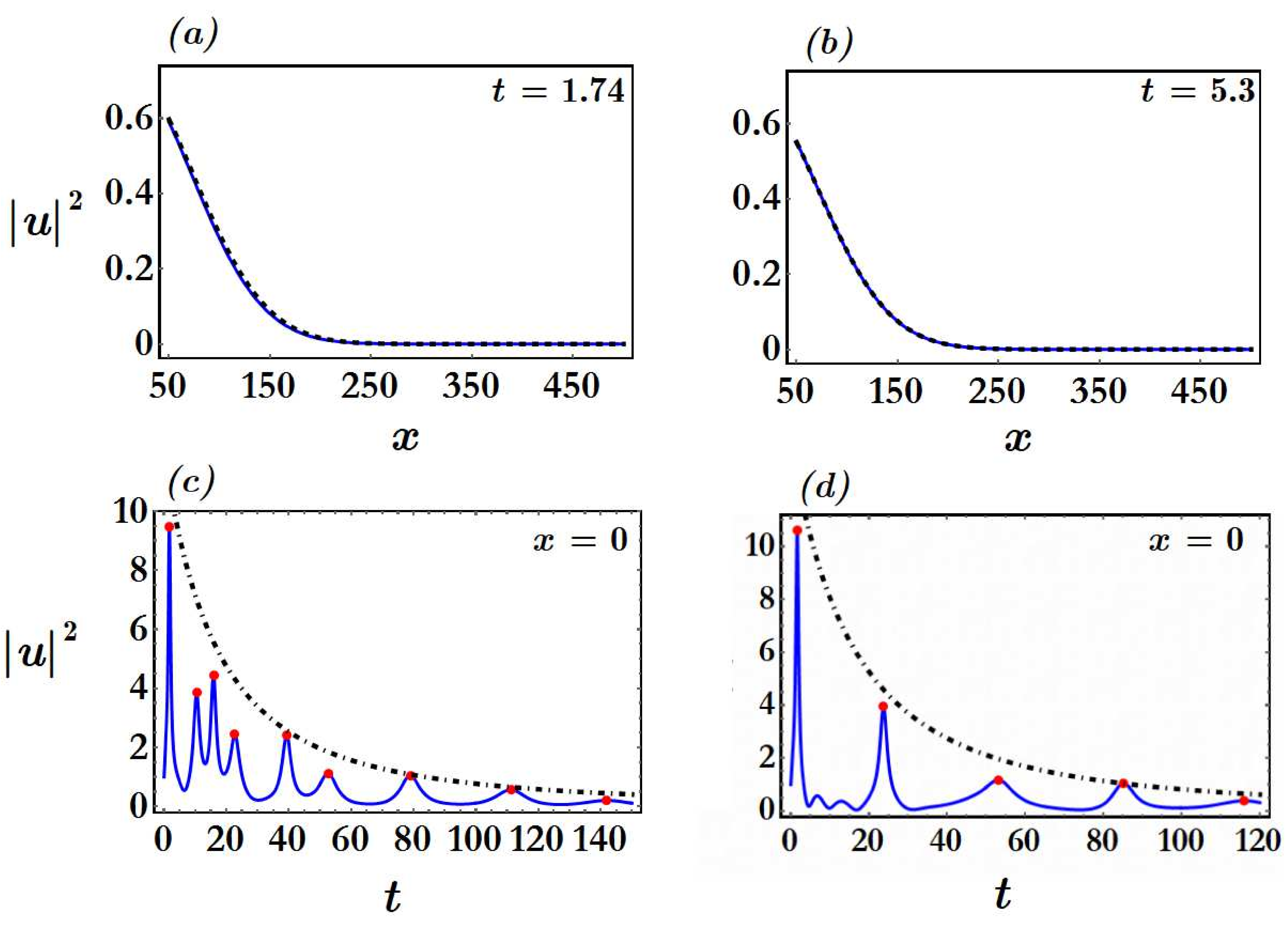}
	\caption{(Color Online) Decaying rates in the presence of the Gaussian driver \eqref{wlf}. Panel (a): Decay of the density of the numerical solution for $x\geq 50$ [continuous (blue) curve], against the decay of the Gaussian function $\rho^{-2}(x)\propto \exp\left[-\left(\frac{x+x_0}{\sigma}\right)^2\right]$, with $x_0\sim 10^{-4}$ and $\sigma=102.3$ [dashed (black) curve], at $t=1.74$. Initial condition \eqref{ic1}, $\gamma=0.01$, $\Gamma=1$, $L=500$, $\sigma_x=100$, $\sigma_t=0.5$. Panel (b): Same as in panel (a), but for $t=5.3$. Panel (c): Temporal decay of the density of the center $|u(0,t)|^2$ of the numerical solution, against the decay of the function $\phi^{-1}(t)\propto \left[1+\frac{(t+s_0)}{t_0}\right]^{-2}$, with $s_0\sim 10$ and $t_0\sim 24.06$ [dotted-dashed (black) curve].  Algebraically decaying initial condition \eqref{ic1}, and the rest of parameters as above. Panel (d): Same as in panel (c) but for $s_0=2.18$, $t_0=29.83$, and the $\mathrm{sech}$-profiled initial condition \eqref{ic2exp}. The rest of parameters are fixed as above. }
	\label{figure3}
\end{figure}

Next, we want to investigate the relevance of the generic spatiotemporal estimates to the observed dynamics. Clearly, the PRW-type event itself, is covered by the algebraic space-time decaying rates predicted by Theorem \ref{spti},  since the PRW solitonic structure possesses algebraic spatiotemporal localization (following closely the analytical PRW), on the top of the decaying support.  However, we also want to study the asymptotics of solutions as $|x|\rightarrow\infty$. For this purpose, we performed a least-square curve fitting of the numerically acquired data, excluding the central PRW event. From the best fitting curve we acquired the decay rate of the tails of the solutions, while by performing small translations (if needed), we derived the corresponding bounding curve. 
	
By applying the above procedure, we revealed that  the tails of the solutions are not uniformly algebraically decaying; instead, the decaying rates of the support  are found to be switched to Gaussian, as shown in Fig. \ref{figure3}. The upper panels (a) and (b) demonstrate the  decay rate of the solutions [continuous (blue) curve], with initial condition \eqref{ic1}, 
for $x\geq 50$, when $t=1.74$ and $t=5.3$,  respectively. It is shown that $|u(x,t)|^2\propto \frac{1}{\rho^2(x)}$, where $\rho^2(x)=\exp\left[\left(\frac{x+x_0}{\sigma}\right)^2\right]$, with $x_0\sim 10^{-4}$ and $\sigma=102.3$ [dashed (black) curve]. We observe that the fitting is almost exact, particularly for $t=5.3$. The same spatial decay rates were detected, also for the  solution with initial condition \eqref{ic2exp} (not shown here). Although the detected spatial decay rates, do not contradict the generic estimates (since the  numerical decay rate is faster), they suggest that the spatial decay rate of the driver seems to be the one which determines the decaying rate of the vanishing support, and not that of the initial condition. 

 On the other hand,  the analytical predictions seem to capture well the time-decay rate of the numerical solution, uniformly in time.  In order to establish this effect, we performed a fitting as described above, on the peaks of the time evolution of the density of the center $|u(0,t)|^2$ of the solutions with initial conditions \eqref{ic1} and \eqref{ic2exp}. The acquired results are shown in  Panels (c) and (d) of Fig.~\ref{figure3} respectively, together with the points used for the fitting (red dots) and the corresponding bounding curve. Panel (c) verifies that $|u(0,t)|^2\propto\frac{1}{\phi(t)}$, where $\phi (t)=\left[1+\frac{(t+s_0)}{t_0}\right]^2$, with $s_0\sim 10$ and $t_0\sim 24.06$ [dotted-dashed (black) curve].  Panel (d) shows that  $|u(0,t)|^2\propto\frac{1}{\phi(t)}$,  but for $s_0=2.18$ and $t_0\sim 29.83$; we observe that the quadratically decaying bounding curve describes very well the decay rate of the peaks of the temporal oscillations.

Summarizing, in the case of the Gaussian forcing we found the following: The spatial algebraic decaying estimates describe the actual decay of the solution in a neighborhood around the PRW-type event, while the sustaining support is decaying with a Gaussian rate, as it is determined by the spatial decaying rate of the driver. Regarding the temporal decay of the solution, it is found that its peaks follow a quadratic decaying rate. 
\begin{figure}[tbh!]
	\hspace{-0.5cm}\includegraphics[scale=0.18]{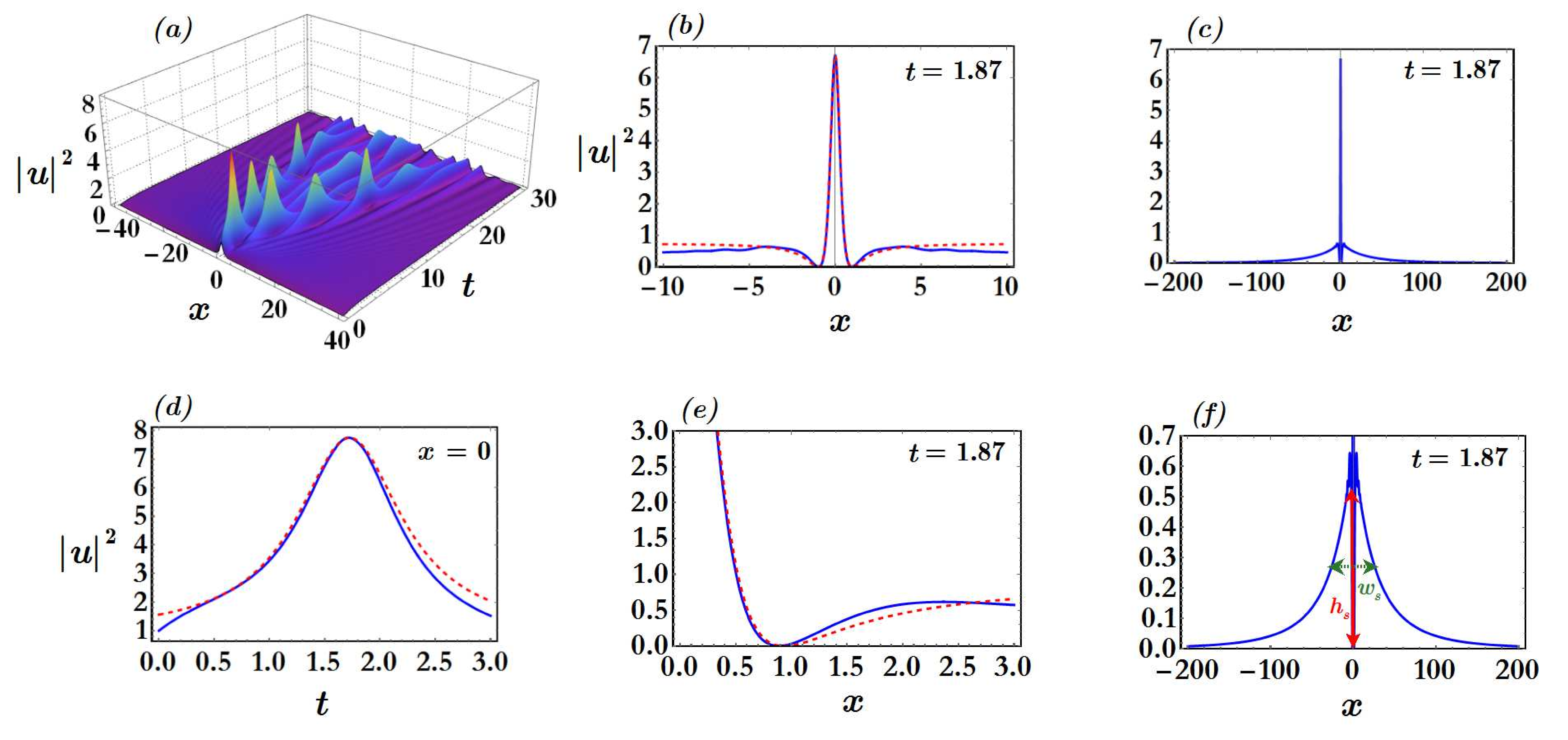}
	\caption{(Color Online) Dynamics of the algebraically decaying initial condition \eqref{ic1}, in the presence of the weakly localized driver \eqref{wlf1}. Parameters $\gamma=0.01$, $\Gamma=1.5$, $\delta_x=100$, $\delta_t=0.5$, $L=400$. Panels (a)-(f) describe the same dynamical features  as in Fig.~\ref{figure1}. The various characteristics of the numerical PRW-type event are plotted against those of the analytical PRW  $u_{\mbox{\tiny PS}}(x,t;1.87;0.745)$. }
	\label{figure4}
\end{figure}

\subsection{Algebraically localized driver}
We continue with the study of the evolution of the vanishing initial conditions \eqref{ic1} and \eqref{ic2exp} (which have been already used in the previous section), but in the presence of the weakly localized driver \eqref{wlf1}. We only present  the results for the quadratic case ($\omega=\theta=0$),  since the corresponding results for the quartic decaying driver ($\omega,\theta>0$) were found almost identical to the quadratic one. 

\begin{figure}[tbh!]
	\hspace{-0.5cm}\includegraphics[scale=0.18]{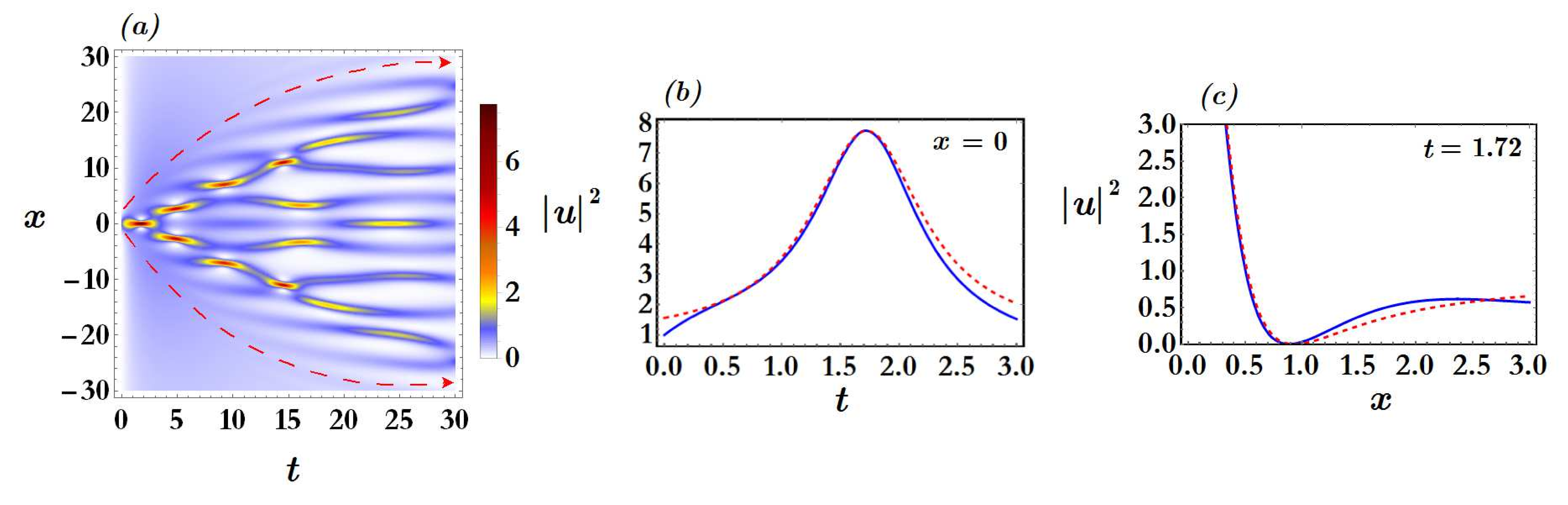}
	\caption{(Color Online) Dynamics of the exponentially decaying initial condition \eqref{ic2exp}, in the presence of the weakly localized driver \eqref{wlf1}. Parameters $\gamma=0.01$, $\Gamma=1.5$, $\delta_x=100$, $\delta_t=0.5$, $L=400$.  Panels (a)-(c) describe the same dynamical features as in Fig.~\ref{figure3}. The  characteristics of the numerical PRW-type event are plotted against those of analytical PRW  $u_{\mbox{\tiny PS}}(x,t;1.72;0.86)$. }
	\label{figure6}
\end{figure}

The emergence of PRW-type solitonic structures, within a reminiscent of semi-classical type dynamics, still persists, however, we may also identify important differences in comparison with the case of the Gaussian driver:  First, by examining the 3D-graphs shown in panels (a) of  Fig.~\ref{figure1} and Fig.~\ref{figure4},  we observe a change in the pattern of the spatiotemporal oscillations; in particular, their density and number within the region bounded by the caustics [traced by the dashed (red) curves shown in panels (a) of Fig.~\ref{figure2} and Fig.\ref{figure6}], is decreased.  Second, by comparing the panels (b) of Fig.~\ref{figure1} and Fig.~\ref{figure4}, we observe, in the present case, a decrease in the amplitude of the PRW-event. Third, in panels (d) and (e) of Fig.~\ref{figure1} and Fig.~\ref{figure4}, we see that, in the previously examined case, the time growth/decay of the PRW-event is closer to the analytical PRW (e.g. they share the same rates for a larger time interval). Fourth, by comparing the panels (c) and (f) of Figs. \ref{figure1} and \ref{figure4},  we observe that the profile of the decaying support has drastically changed, suggesting accordingly, for a change of its decaying rate. 

\begin{figure}[tbh!]
	\hspace{-0.5cm}\includegraphics[scale=0.18]{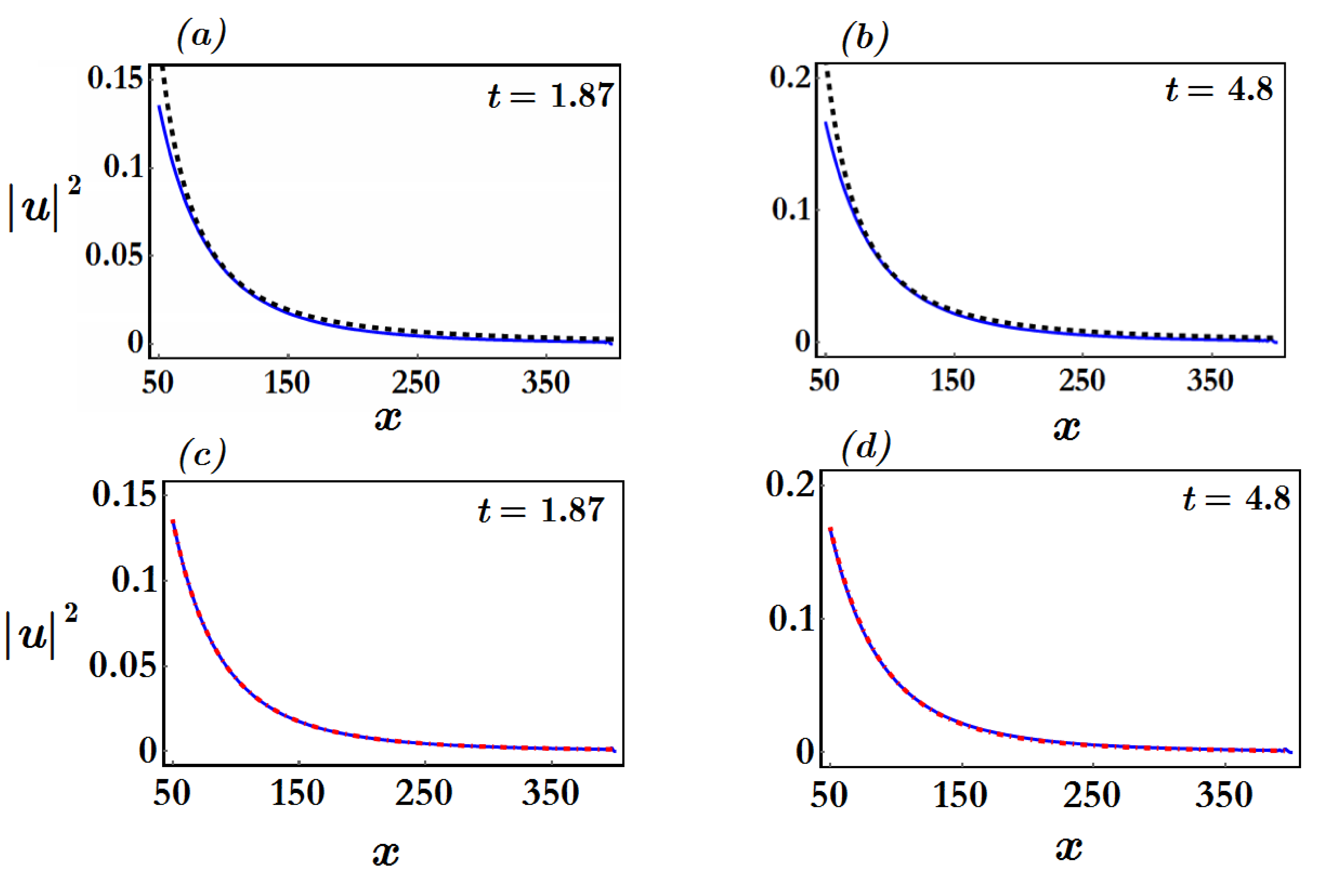}
	\caption{(Color Online)  Decaying estimates in the presence of the weakly localized driver \eqref{wlf1}. Panel (a): Decay of the density of the numerical solution for $x\geq 50$ [continuous (blue) curve], against the decay of the function $\rho^{-2}(x)\propto \left(1+\left|\frac{x}{x_0}\right|\right)^{-2}$, with $x_0\sim 5.95$ [dashed (black) curve],  at $t=1.87$.  Algebraically decaying initial condition \eqref{ic1}, $\gamma=0.01$, $\Gamma=1$, $L=400$, $\delta_x=100$, $\delta_t=0.5$. Panel (b): Same comparison as in panel (a), but for $t=4.8$. Panel (c): Same as above, but for $\rho^{-2}(x)\propto\left(1+\left|\frac{x}{x_0}\right|+\frac{x^2}{x_0^2}\right)^{-2}$ with $x_0\sim 100$. Panel (d): Same as in panel (c), but for $t=4.8$. }
	\label{figure7}
\end{figure}

In order to highlight the differences in the spatial decay rates for the two types of the driver, we proceed with a study on the tails of the solutions in the algebraic case.
Fig.~\ref{figure7}
illustrates that the spatial decay of the solution is uniformly algebraic. The upper panels (a) and (b) show the bounding curve of the  density of the solution acquired from the fitting process  [continuous (blue) curve], with the algebraically decaying initial condition \eqref{ic1}, for $x\geq 50$, when $t=1.87$ and $t=4.8$,  respectively. We observe that $|u(x,t)|^2\propto \frac{1}{\rho^2(x)}$, where $\rho^2(x)=\left(1+\left|\frac{x}{x_0}\right|\right)^2$, with $x_0\sim 5.95$ [dashed (black) curve], i.e., when the weight example \eqref{eqn6} is used. This approximation captures with a very good agreement the decay rate of the tail of the support for $x\geq 70$.  The algebraic approximation of the spatial decay becomes almost exact for a quadratic correction of the weight $\rho(x)$: Panels (c) and (d) depict the above comparison,  but for $\rho^2(x)=\left(1+\left|\frac{x}{x_0}\right|+\frac{x^2}{x_0^2}\right)^2$ with $x_0\sim 100$.  The spatially uniform decay rates are also preserved in the case of the exponential decaying initial condition \eqref{ic2exp} (not shown here). 

\begin{figure}[tbh!]
	\hspace{-0.5cm}\includegraphics[scale=0.18]{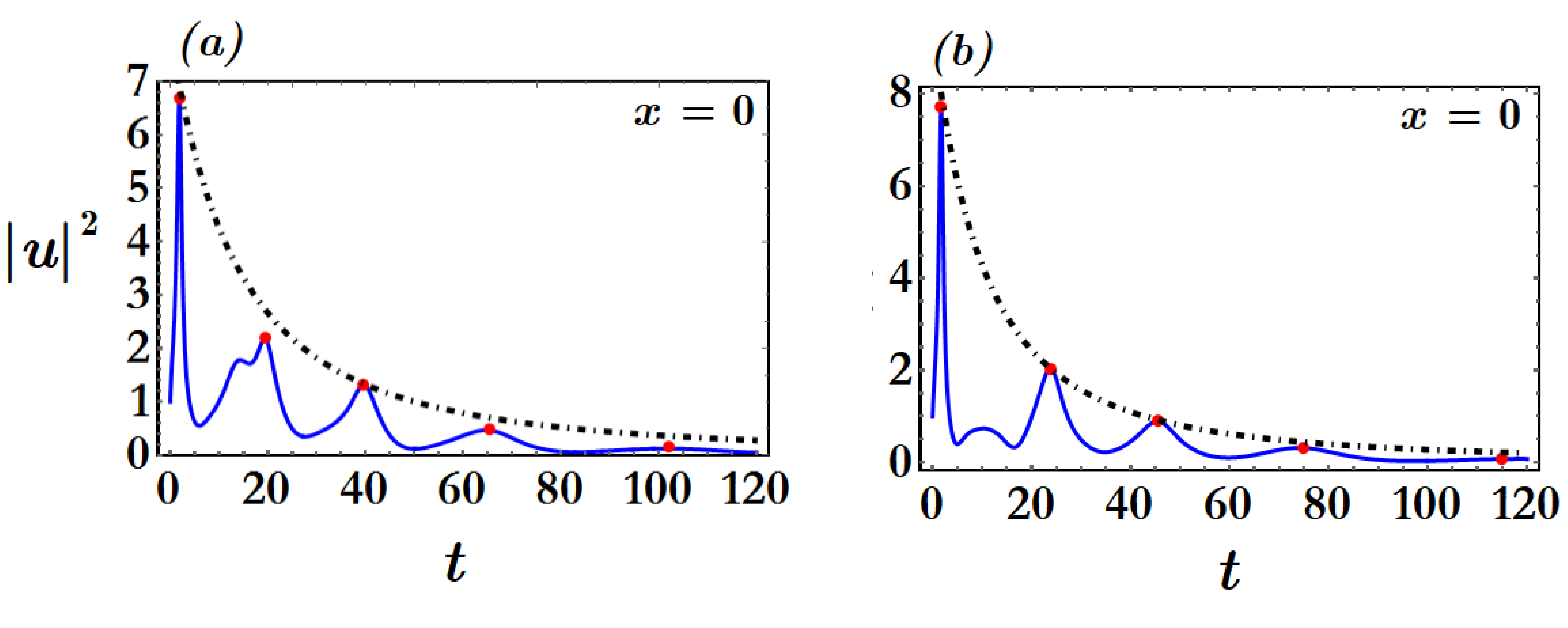}
	\caption{(Color Online) Panel (a): Temporal decay of the density of the center $|u(0,t)|^2$ of the numerical solution, against the decay of the function $\phi^{-1}(t)\propto \left[1+\frac{(t+s_0)}{t_0}\right]^{-2}$, with $s_0\sim 15.7$ and $t_0\sim 12$ [dotted-dashed (black) curve].  Algebraically decaying initial condition \eqref{ic1} and the rest of parameters as in Fig.~\ref{figure7}. Panel (b): Same as in panel (c) but for $s_0=0$, $t_0=20.75$, and the $\mathrm{sech}$-profiled initial condition \eqref{ic2exp}. The rest of parameters are fixed as in Fig.~\ref{figure7}. }
	\label{figure8}
\end{figure}

The analytically derived temporal decay rates, are proved  to be very relevant in describing the decay of the numerical solution with respect to time, as justified in Fig.~\ref{figure8}: Panel (a) portrays the decay of the density of the center $|u(0,t)|^2$, for the numerical solution starting from the algebraically decaying initial condition \eqref{ic1}. The plot verifies that $|u(0,t)|^2\propto\frac{1}{\phi(t)}$, where $\phi (t)=\left[1+\frac{(t+s_0)}{t_0}\right]^2$, with $s_0\sim 15.7$ and $t_0\sim 12$ [dotted-dashed (black) curve]. Panel (b) depicts the same comparison for the exponentially decaying initial condition \eqref{ic2exp}; the density of the center decays as $|u(0,t)|^2\propto \left(\frac{1}{\phi(t)}\right)$, with $s_0=0$ and $t_0=20.75$. In both cases of initial conditions, the fitting curves describe in a very good agreement the decay rate of the peaks of the temporal oscillations. 

As a summary, in the case of the weakly localized forcing, we found that the solution decays both spatially and temporarily with algebraic rates, uniformly, as predicted by the analytical estimates. In particular, for the spatial decay, the actual decay rates are not only justified by the analytical ones in the neighborhood of the emerged PRW, but predict accurately for large $|x|$, the decaying rate of the vanishing support.  At the same time, the analytically predicted decaying rates, were still found to be in a very good agreement with the numerical ones, in describing the (decreasing in amplitude) time-oscillations of the solution.
\begin{figure}[tbp]
	\begin{center}
		\includegraphics[scale=0.252]{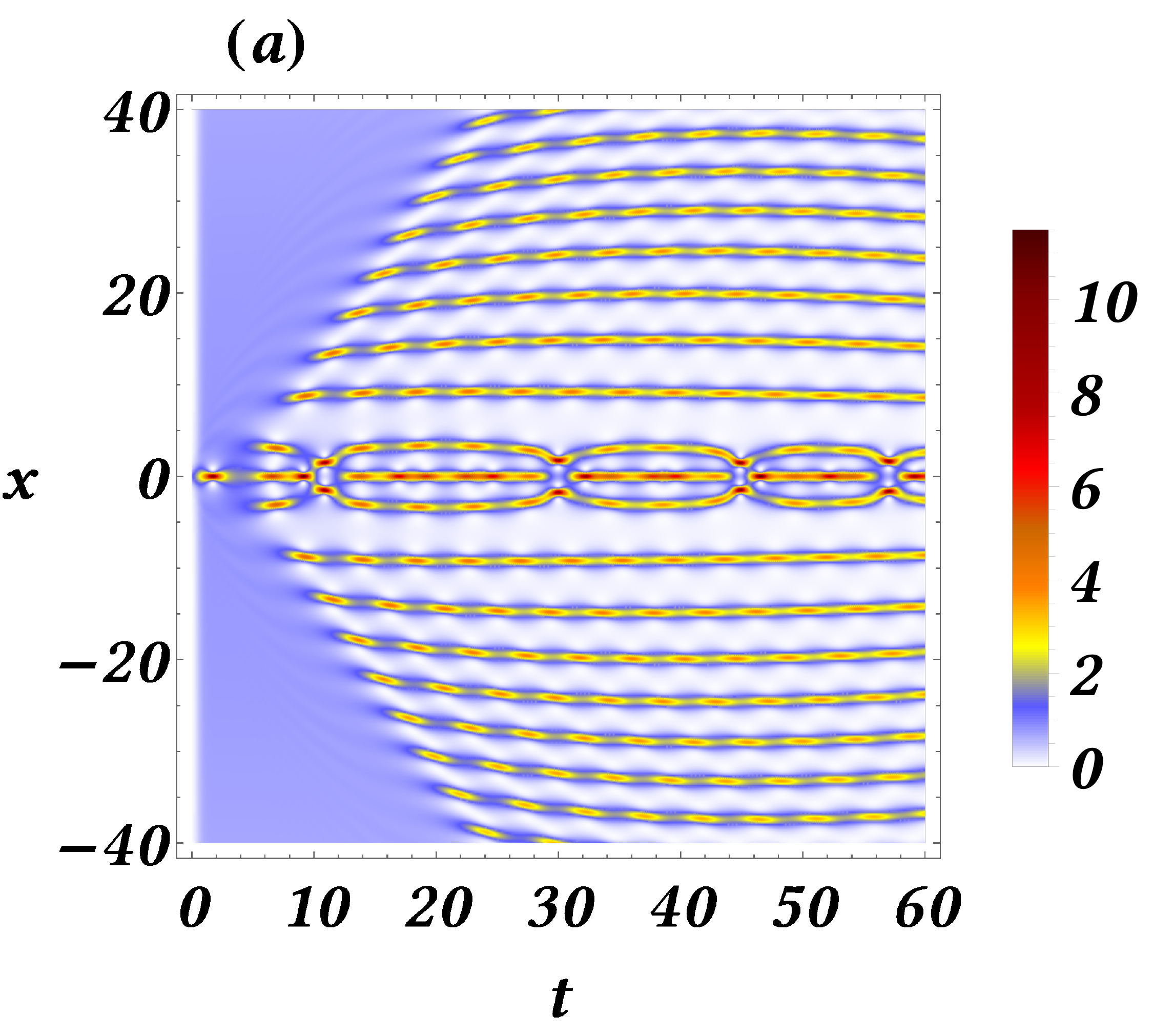}
		\includegraphics[scale=0.252]{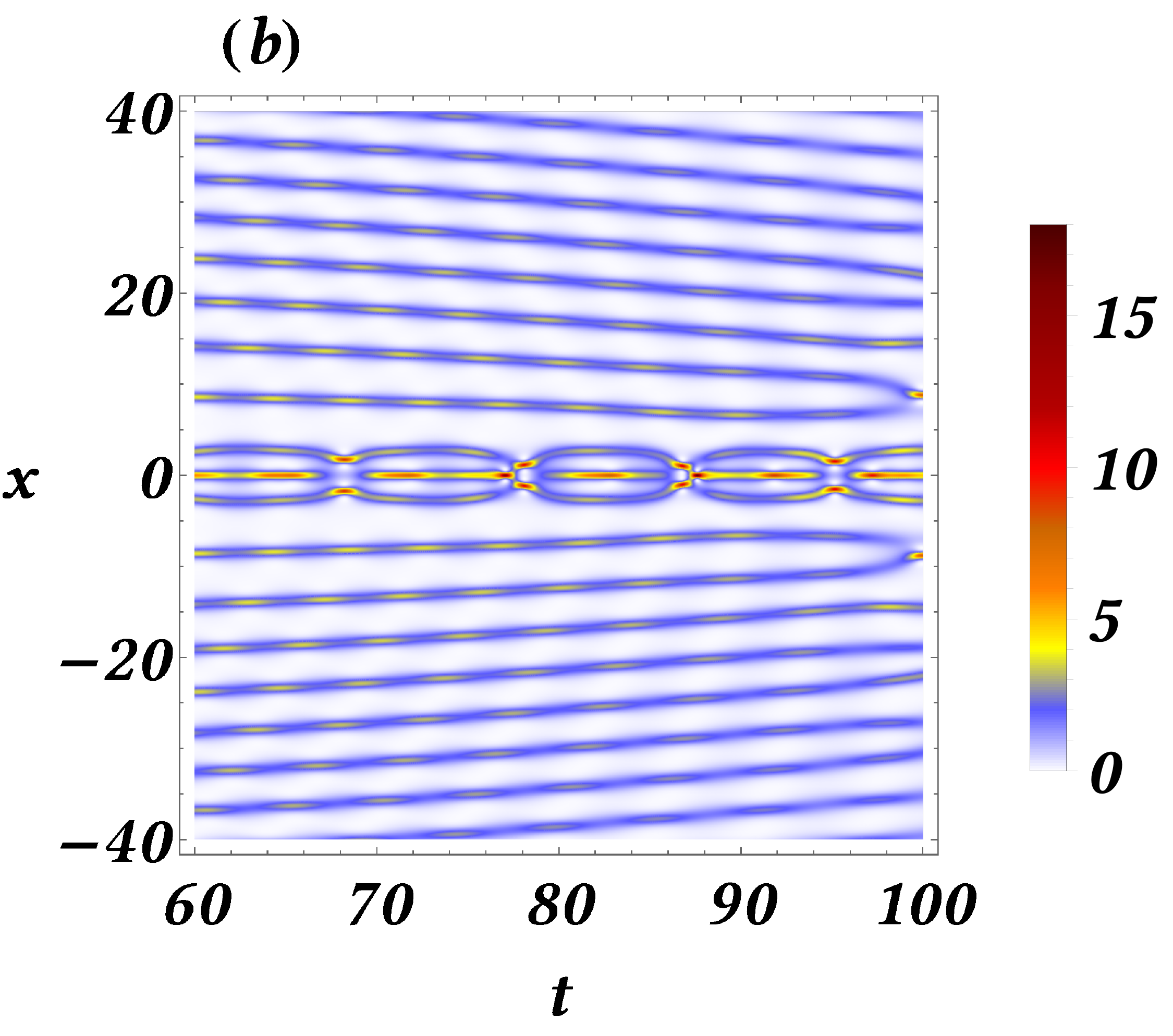}
		\includegraphics[scale=0.252]{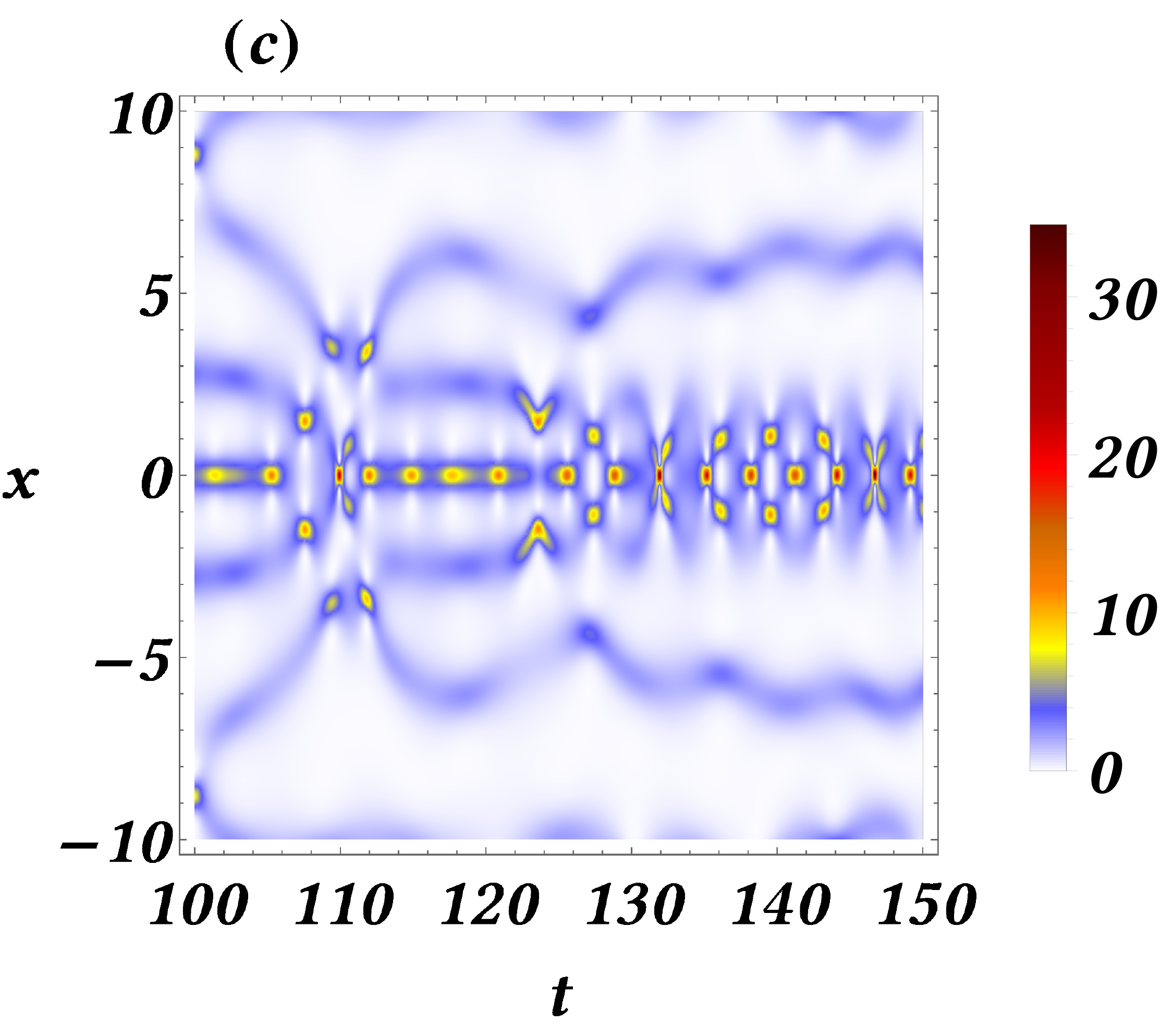}		
	\end{center}
	\caption{(Color Online) Contour plots of the spatiotemporal evolution of the density in the undamped and forced limit $\gamma=0$, $\Gamma=1$ for the quadratically decaying initial condition \eqref{ic1}. The rest of parameters are fixed as in the study of Figure \ref{figure1}. Panel (a): $t\in [0,60]$. Panel (b): $t\in [60,100]$. Panel (c): $t\in [100,150]$.}
	\label{figure8N}
\end{figure}	
\subsection{Comments in the case of the undamped and driven limit ($\gamma=0$, $\Gamma\neq 0$)}
In the case of the undamped but driven limit ($\gamma=0$, $\Gamma\neq 0$), the analytical arguments, on the uniform-in-time estimates in the weighted norms, are not valid. Thus, we are unable to acquire estimates on the the spatiotemporal decaying rates established in the framework of Propositions \ref{spatial} and \ref{P1}.

Despite the lack of analytic estimates, in order to study the dynamics of the system in this limit, we perform a numerical study for $\gamma=0$, $\Gamma=1$, while the other parameters are fixed as in the study of Figure \ref{figure1}, for the quadratically decaying initial condition \eqref{ic1}. The results, for this parametric regime, are shown in Figure \ref{figure8N}, depicting the spatiotemporal evolution of the density for the time intervals $[0,60]$ (panel (a)), $[60,100]$ (panel (b)) and $[100,150]$ (panel (c)), respectively. In panel (a), we observe that while the emergence of the PRW-event yet persists in this limit, a drastic change appears in the spatiotemporal pattern if compared with the one shown panel (a) of Figure \ref{figure2}. Due to the presence of solely the driver, the "lattice" of extreme events (in the form of PRW-structures) which was partially reminiscent of the semiclassical NLS type dynamics of \cite{BM1},\cite{BM2} is now replaced by an ``array" of breathing modes.  We also observe the almost  periodic pattern around the center, consisting of a centralized, "hybrid" solitonic/breathing mode surrounded by breathing-like modes which collide elastically to form a chain-like pattern. Progressively, as shown in panel (b), the paths of the breathers forming the array are focusing towards the central structure which gradually increases its amplitude. Finally (panel (c)), a big percentage of the energy of the surrounding modes is also concentrated to the the central one, forming this way a dominating centralized breathing mode possessing increasing amplitude.

Thus, in contrast with the damped case $\gamma>0$, the above dynamical picture suggests that the solutions become unbounded in the absence of the damping, despite the fact that the spatiotemporally localized driver is asymptotically vanishing. 
\section{Conclusions}\label{SIV} In this work, we have proved spatiotemporal algebraically decaying estimates for the density of the solutions of the linearly damped nonlinear Schr\"odinger equation, in the presence of  localized driving forces. The problem is endowed with vanishing boundary conditions in the real line.  The analytical results on the spatial decaying rates were derived through an approximation scheme, which corroborates estimates in weighted Sobolev norms for an auxiliary problem (supplemented with Dirichlet boundary conditions), in order to avoid restrictive a priori radiation assumptions. Then, the passage to the limit, takes advantage of the independence of the constants of the usual interpolation inequalities from the spatial interval. For the derivation of the temporal decaying estimates, we implemented in the simplified one-dimensional set-up, a time-weighted energy method which originates from the analysis of similar questions for the solutions of the Navier-Stokes equation.  

The numerical simulations revealed that the analytically derived decaying rates proved particularly relevant, as an attempt to rationalize the transient dynamics of the model: on the one hand, they capture the dynamics of extreme wave events (strongly reminiscent to the Peregrine rogue wave of the integrable NLS), as spatiotemporal algebraically localized wave forms. On the other hand, it was found that they describe, in most cases with a very good agreement, the space-time asymptotics of the numerical solutions. 

Future extensions, may consider the study of the model in higher-spatial dimensions. Such a task, may involve interesting implications induced by the functional-analytic limitations of the phase spaces in dealing with the potential criticality of the nonlinearity exponents. Another direction, concerns the study of the model when supplemented with periodic boundary conditions. Such a consideration is of particular physical interest as it includes the case of initial data vanishing on a finite background; it may non-trivially involve the impact of interpolation inequalities for periodic functions, and the various dependencies of their optimal constants. It is also certainly relevant to study discrete counterparts in multidimensional lattices, where the interplay of discreteness, dimensionality and nonlinearity may give rise to fascinating dynamics \cite{KevreDNLS},\cite{CNSNS2018}. Relevant investigations are in progress, and outcomes will be reported in upcoming publications.
\section*{Acknowledgments}
The authors acknowledge that this work was made possible by the NPRP grant \# [8-764-160] from the Qatar National Research Fund (a member of Qatar Foundation). The findings achieved herein are solely the responsibility of the authors.

\end{document}